\documentclass[11pt]{article}
\input{psfig.sty}
\begin{document}
%%%%%%%%%%%%%%%%%%%%%%%%%%%%%%%%%%%%%%%%%%%%%%%%%%%%%%%%%%%%%%%%%%%

\title{
\textbf{Transfer matrix and Monte Carlo tests of critical exponents
in Ising model}
}

\author{J. Kaupu\v{z}s
\thanks{E--mail: \texttt{kaupuzs@latnet.lv}} \\
Institute of Mathematics and Computer Science, University of Latvia\\
29 Rainja Boulevard, LV--1459 Riga, Latvia}

\date{\today}

\maketitle

\begin{abstract}
The corrections to finite-size scaling in the critical two-point correlation
function $G(r)$ of 2D Ising model on a square
lattice have been studied numerically by means of exact transfer--matrix
algorithms. The systems
have been considered, including up to 800 spins. The calculation of $G(r)$
at a distance $r$ equal to
the half of the system size $L$ shows the existence of an amplitude
correction $\propto L^{-2}$. A nontrivial correction $\propto L^{-0.25}$
of a very small magnitude also has been detected, as it
can be expected from our recently developed GFD
(grouping of Feynman diagrams) theory.
Monte Carlo simulations of the squared magnetization 
of 3D Ising model have been performed by Wolff's algorithm in the
range of the
reduced temperatures $t \ge 0.000086$ and system sizes $L \le 410$.
The effective critical exponent $\beta_{eff}(t)$
tends to increase above the currently accepted numerical values.
The critical coupling $K_c=0.22165386(51)$ has been extracted from the 
Binder cumulant data within $L \in [96;384]$. 
The critical exponent $1/\nu$, 
estimated from the finite--size scaling of the derivatives of the Binder 
cumulant, tends to decrease slightly below the RG value $1.587$
for the largest system sizes. 
The finite--size scaling of accurately simulated
maximal values of the specific heat $C_V$ in 3D Ising model
confirms a logarithmic rather than power--like critical
singularity of $C_V$.

\end{abstract}


\vspace*{1ex}

{\bf Pacs:} 64.60.Cn, 68.18.Jk, 05.10.-a

\section{Introduction}

Since the exact solution of two--dimensional Lenz--Ising (or Ising)
model has been found by Onsager~\cite{Onsager}, a study of various phase
transition models is of permanent interest. Nowadays, phase
transitions and critical phenomena is one of the most widely
investigated fields of physics~\cite{Sornette,BDT}.
Remarkable progress has been
reached in exact solution of two--dimensional models~\cite{Baxter}.
Recently, we have proposed~\cite{K1} a novel method based on
grouping of Feynman diagrams (GFD) in $\varphi^4$ model.
Our GFD theory allows to analyze the asymptotic solution
for the two--point correlation function at and near criticality,
not cutting the perturbation series. As a result the possible
values of exact critical exponents have been proposed~\cite{K1} for
the Ginzburg--Landau ($\varphi^4$) model with $O(n)$ symmetry,
where $n=1, 2, 3, \ldots$ is the dimensionality of the order
parameter. Our predictions completely agree with the
known exact and rigorous results in two dimensions~\cite{Baxter}.
In~\cite{K1},
we have compared our results to some Monte Carlo (MC) simulations
and experiments~\cite{IS,SM,GA}. The examples considered there
support our predictions about the critical exponents. A more recent comparison with 
experimental data very close to the $\lambda$--transition point 
$T=T_{\lambda}$ in liquid helium has been made in~\cite{Khe}. 
We have shown there that our critical exponents better describe 
the closest to $T_{\lambda}$ data 
for the superfluid fraction of liquid helium as compared to  
%It has been shown~\cite{K1} that the
%actually discussed MC data for 3D Ising~\cite{IS} and
%$XY$~\cite{SM} models are fully consistent with our theoretical
%predictions, but not with 
the exponents provided by the perturbative renormalization
group (RG) theory~\cite{Wilson,Ma,Justin}.
As claimed in~\cite{K1}, the conventional RG expansions
are not valid from the mathematical point of view.
The current paper, dealing with numerical transfer-matrix analysis of
the two--point correlation function in 2D Ising model, as well as with 
MC simulations in the three--dimensional Ising model
presents some more evidences
in favour of the critical exponents predicted by the GFD theory.
Our estimations are based on the finite--size scaling theory,
which by itself is an attractive field of investigations~\cite{Chamati}
and has increasing importance in modern physics~\cite{BDT}.

\section{Critical exponents predicted by GFD theory} \label{sec:crex}

 Our theory predicts possible values of exact critical exponents
$\gamma$ and $\nu$ for the $\varphi^4$ model whith $O(n)$
symmetry ($n$--component vector model) given by the Hamiltonian
\begin{equation} \label{eq:Ha}
H/T= \int \left[ r_0 \varphi^2({\bf x})
+ c (\nabla \varphi({\bf x}))^2
+ u \varphi^4({\bf x}) \right] d{\bf x} \; ,
\end{equation}
where $r_0$ is the only parameter depending on temperature $T$,
and the dependence is linear.  At the spatial
dimensionality $d=2, 3$ and $n=1, 2, 3, ...$ the predicted
possible values of the critical exponents are~\cite{K1}
\begin{eqnarray} 
\gamma &=& \frac{d+2j+4m}{d(1+m+j)-2j} \label{eq:gamma}
\label{eq:expo1} \; , \\
\nu &=& \frac{2(1+m)+j}{d(1+m+j)-2j} \label{eq:nu}
\label{eq:expo2} \; ,
\end{eqnarray}
where $m \ge 1$ and $j \ge -m$ are integers. It is well known that
the $O(n)$--symmetric $\varphi^4$ model belongs to the same universality
class as the corresponding lattice model (Ising model at $n=1$, $XY$
model at $n=2$, the classical Heisenberg model at $n=3$, etc.), where
the order parameter is an $n$-component vector (spin) with fixed modulus
$\mid \varphi({\bf x}) \mid=1$, since the latter is a particular case
($r_0 \to -\infty$ at $-r_0/(2u)=1$ or $\lambda \to \infty$ in the
notations used in~\cite{Has}) of the lattice $\varphi^4$ model, where
the gradient term is represented by finite differences~\cite{Has}.
Besides, the partition functions and two--point correlation functions of
both $\varphi^4$ model in~\cite{K1} and Ising model can be represented
by similar functional integrals~\cite{Amit,Fproc}. Thus, at $n=1$ we
have $m=3$ and $j=0$ to fit the known exact results for the
two--dimensional Ising model. As proposed in Ref.~\cite{K1}, in the case
of $n=2$ we have $m=3$ and $j=1$, which yields in three dimensions
$\nu=9/13$ and $\gamma=17/13$.
%%%%%%%%%%%%%%%%%%%%%%%%%%%%%%%%%%%%%%%%%%%%%%%%%%%%%% 

As already
explained in~\cite{K1}, our predictions do not refer to the case of the
self--avoiding random walk recovered at $n=0$. The
values~(\ref{eq:gamma}) and~(\ref{eq:nu}) have been derived in~\cite{K1}
assuming that $2 \nu -\gamma>0$ holds. In principle, the
mean--field--like solution with $2 \nu- \gamma=0$ can exist at $d<4$,
and it refers to the Gaussian random walk with $n=-2$. This is a special
case, not related to~(\ref{eq:gamma}) and~(\ref{eq:nu}), where two
expansion parameters $\Delta^{2 \nu -\gamma}$ and $\Delta^{2 \gamma- d
\nu}$ with $\Delta=T-T_c \to 0$ being the deviation from the critical
temperature $T_c$, are replaced by one parameter $\Delta^{2 \gamma- d
\nu}$. Eq.~(48) in~\cite{K1}  can be then satisfied with $\gamma=1$ and
$\nu=1/2$, i.~e., all the exponents are consistent and each term can be
compensated. The singularity of the specific heat with the exponent
$\alpha=2-d/2$ comes from the leading terms in Eq.~(60) of~\cite{K1}.
Obviously, $\gamma=1$ and $\nu=1/2$ always are the true exponents at
$d>4$, where the Gaussian approximation  $G({\bf k})= 1/ \left[G({\bf
0})^{-1} + 2ck^2 \right]$ for the two-point correlation function $G({\bf
k})$ in the Fourier representation is asymptotically exact at $u \to 0$
and $T>T_c$ for arbitrarily small wave vectors ${\bf k}$. These
exponents are recovered at any $m$ and $j$ in~(\ref{eq:gamma})
and~(\ref{eq:nu}) when approaching the upper critical dimension $d=4$
from below. 

Our formulae do not provide any sensible result approaching
$d=1$, where $\nu=1/(d-1)$ is expected at $n=1$ according to the
Midgal's approximation~\cite{Stephen}.  It can be understood from the
point of view~\cite{Kst} that $2$,  probably, is the marginal value of
$d$, such that an analytic continuation  of the results from
$d$-dimensional hypercubes can be only formal and  has no  physical
meaning at $d<2$.  In this sense, we
expect that $d=2$ is a special dimension for any $n \ge 1$. Besides, the
critical temperature does not vanish at $d \to 2+0$, and for $n=1$ there
exist  lattices for  which the critical temperature is nonzero at
the fractal dimension  below 2~\cite{ReMa}. 
In the marginal case $d=2$ different behavior is observed
at low temperatures: the long--range order at $n=1$, the
Kosterlitz--Thouless structural order at $n=2$, and disordered
state at $n>2$. 

Our concept agrees with the known rigorous results
for $XY$ model~\cite{TKHT,FS}. It disagrees with the prediction
of the perturbative RG theory~\cite{BZ} that the critical temperature
goes to zero at $d \to 2+0$ for the $O(n>2)$--symmetric nonlinear
$\sigma$ model and, therefore, the behavior in this case is Gaussian,
i.~e., $\eta=0$ and $\nu=1/(d-2)$.
The results of the perturbative RG theory are not rigorous since
the claims are based on formal expansions which break down
in relevant limits, in this case at vanishing external field $H \to +0$.
Moreover, essential claims of this theory are based
on an evidently incorrect mathematical treatment.
In particular, the conclusion about
the Gaussian character of the $O(n)$--symmetric $\varphi^4$ model
below $T_c$ has been made in~\cite{Law} by simply rewritting
the Hamiltonian in an apparently Gaussian form
(see Eqs.~(3.4) to~(3.6) in~\cite{Law}). The author, however,
forgot to include the determinant of the transformation Jacobian
in the relevant functional integrals, according to which the resulting
model all the same is not Gaussian.
Due to the reasons mentioned
above, we do not believe in predictions of the perturbative RG
theory, but rely only on exact and rigorous results.

 There exists a simple non--perturbative explanation why the critical
temperature should stay finite at $d \to 2+0$ for the
$O(n)$--symmetric Heisenberg model. Below $T_c$, the difference in
free energies for models
with antiperiodic and periodic boundary conditions along one axis
is $\Delta F \propto \Upsilon(T) L^{d-2}$, where $L$ is the linear
size of the system and $\Upsilon(T)$ is the helicity modulus.
It holds because the energy difference in the ground state at $T=0$ is
$\propto L^{d-2}$, corresponding to gradually rotated spins in any given
plane. The factor $\Upsilon(T)$ takes into account the
temperature dependence. It vanishes at $T \ge T_c$.
Hence, the factor $L^{d-2}$ always vanishes at $d<2$ in the thermodynamic
limit $L \to \infty$, therefore the long--range order (if it would exist) 
could be destroyed in this case at any finite temperature by gradually
rotating the spins without increasing the free energy. Thus, the long--range
order disappears at $d<2$ irrespective to the behavior of $\Upsilon(T)$,
i.~e., irrespective to the value of $T_c$ at $d=2+0$.
In such a way, the assumption that the critical temperature should
go to zero continuously appears as an additional unnecessary constraint.
On the other hand, if the critical temperature remains finite
in $\varphi^4$ model, then $\eta$ should be
positive at $d = 2+0$ to avoid the divergence of
$\langle \varphi^2({\bf x}) \rangle = n \, L^{-d} \sum_{\bf k} G({\bf k})$
at $T=T_c$, where $G({\bf k}) \simeq a \, k^{-2+\eta}$ with
$a \ne 0$ holds for the two--point correlation function.
The expectation $\eta >0$ agrees
with~(\ref{eq:expo1}) and~(\ref{eq:expo2}).
However, the critical temperature at $d=2+0$ and $\eta$ tend to zero 
in the limit $n \to \infty$
to coincide with the known exact results for the spherical model, which
are recovered in~(\ref{eq:expo1}) and~(\ref{eq:expo2}) at $j/m \to \infty$.
%%%%%%%%%%%%%%%%%%%%%%%%%%%%%%%%%%%%%%%%%%%%%%%%%%%%%%%%

In the present analysis the correction--to--scaling
exponent $\theta$ for the susceptibility is also relevant. The susceptibility
is related to the correlation function 
$G({\bf k})$ via $\chi \propto G({\bf 0})$~\cite{Ma}. In the
thermodynamic limit, this relation makes sense at $T > T_c$.
According to our theory, $t^{\gamma} G({\bf 0})$ can be expanded in a Taylor
series of $t^{2 \nu -\gamma}$ at $t \to 0$.
In this case the reduced temperature $t$ is defined as
$t=r_0(T)-r_0(T_c) \propto T-T_c$.
Formally, $t^{2 \gamma - d \nu}$ appears as second expansion
parameter in the derivations in Ref.~\cite{K1}, but,
according to the final result represented by
Eqs.~(\ref{eq:gamma}) and~(\ref{eq:nu}),
$(2 \gamma - d \nu)/(2 \nu -\gamma)$ is a natural number.
Some of the expansion coefficients can be zero, so that in general we have
\begin{equation} \label{eq:Delta}
\theta=\ell \, (2 \nu -\gamma) \; ,
\end{equation}
where $\ell$ may have integer values 1, 2, 3, etc. One can expect
that $\ell=4$ holds at $n=1$ (which yields $\theta=1$ at $d=2$ and
$\theta=1/3$ at $d=3$) and the only nonvanishing
corrections are those of the order $t^{\theta}$, $t^{2 \theta}$,
$t^{3 \theta}$, since the known corrections to scaling for
physical quantities, such as magnetization or correlation length,
are analytical in the case of the two--dimensional Ising model.
Here we suppose that the confluent corrections become analytical,
i.~e. $\theta$ takes the value $1$, at $d=2$.
Besides, similar corrections to scaling are expected for
susceptibility $\chi$ and magnetization $M$ since both these
quantities are related to $G({\bf 0})$, i.~e.,
$\chi \propto G({\bf 0})$ and $M^2=\lim_{x \to \infty}
\langle \varphi({\bf 0}) \varphi({\bf x}) \rangle
= \lim_{V \to \infty} G({\bf 0})/V$
hold where $V=L^d$ is the volume and $L$ is the linear size of
the system. The above limit is meaningful at $L \to \infty$,
but $G({\bf 0})/V$ may be considered as a definition of $M^2$
for finite systems too. The latter means that corrections
to finite--size scaling for $\chi$ and $M$ are similar at $T=T_c$.
According to the scaling hypothesis and finite--size scaling
theory, the same is true for the discussed here corrections at $t \to 0$,
where in both cases ($\chi$ and $M$) the definition
$t= \mid r_0(T)-r_0(T_c) \mid$ is valid.
Thus, the expected expansion of the susceptibility $\chi$ looks
like $\chi = t^{-\gamma} \left( a_0+a_1 t^{\theta} +a_2 t^{2 \theta}
+ \cdots \right)$.
In this discussion we have omitted the irrelevant for critical behaviour background 
term in the susceptibility, which is constant in the first approximation 
and comes from the short--distance contribution to 
$\chi = \sum_{\bf x} G({\bf x})$~\cite{Ma}, where $G({\bf x})$ is the 
real--space two--point correlation function.

Our hypothesis is that $j=j(n)$ and $\ell=\ell(n)$
monotoneously increase with $n$ to fit the known exponents
for the spherical model at $n \to \infty$.
The analysis of the MC and experimental results here and in~\cite{K1}
enables us to propose that $j(n)=n-1$,
$\ell(n)=n+3$, and $m=3$ hold at least at $n=1,2,3$. These relations,
probably, are true also at $n >3$.
This general hypothesis is consistent with the idea that
the critical exponents $\gamma$, $\nu$, and $\theta$
can be represented by some analytical functions of $n$ which are
valid for all natural positive $n$ and yield
$\eta=2-\gamma/\nu \propto 1/n$ rather than $\eta \propto 1/n^s$
with $s=2,3, \ldots$ ($s$ must be a natural number to avoid a
contradiction, i.~e., irrational values of $j(n)$ at natural $n$) at
$n \to \infty$. At these conditions, $j(n)$ and $\ell(n)$ are
linear functions of $n$ (with integer coefficients) such that
$\ell(n)/j(n) \to 1$ at $n \to \infty$, and $m$ is constant.
Besides, $j(1)=0$, $m(1)=3$, and $\ell(1)=4$ hold to coincide with the
known results at $n=1$.
Then, our specific choice is the best one among few possibilities
providing more or less reasonable agreement with the actually discussed
numerical an experimental results.

We allow that different $\ell$  values correspond to
the leading correction--to--scaling exponent for different
quantities related to $G({\bf k})$. According to~\cite{K1}, the expansion of
$G({\bf k})$ in $\varphi^4$ model by itself contains a nonvanishing
term of order
$t^{2 \nu -\gamma} \equiv t^{\eta \nu}$ (in the form
$G({\bf k}) \simeq t^{-\gamma} \left[ g({\bf k} t^{-\nu})
+ t^{\eta \nu} g_1({\bf k} t^{-\nu}) \right]$ whith
$g_1({\bf 0})=0$, since $\ell >1$ holds in the case of susceptibility)
to compensate the corresponding correction term (produced
by $c \left( \nabla \varphi \right)^2$) in the equation
for $1/G({\bf k})$ (cf.~\cite{K1}).

The correction $t^{\eta \nu}$ is related to correction
term $x^{-\eta}$ in the long--distance ($x \to \infty$)
behavior of the real--space pair correlation function
$G(x) \propto x^{2-d-\eta} \left[ 1 + \mathcal{O}
\left( x^{-\eta} \right) \right]$ at the critical point,
as well as to correction $L^{-\eta}$
in the finite--size scaling expressions at criticality.
Such kind of corrections must not necessarily
appear in the Ising model, where they could have zero amplitude.
In particular, the critical Green's (correlation) function $G(x)$
of 2D Ising model in $\langle 11 \rangle$ crystallographic direction on
an infinite lattice can be calculated easily based on the
known exact formulae~\cite{YP}, and
it yields $G(x) \propto x^{-1/4} \left[ 1 + \mathcal{O}
\left( x^{-2} \right) \right]$ at large distances $x \to \infty$.
Nevertheless, our calculations in 2D~Ising model discussed in
Sec.~\ref{sec:result} indicate the existence of a nontrivial
finite--size correction of the kind $L^{-\eta}$
(for $\langle 10 \rangle$ direction),
as it can be expected from our theoretical results for the
$\varphi^4$ model.
The thermodynamic limit is a particular case of the
finite--size scaling with the scaling argument $x/L \to 0$,
therefore it is possible that the nontrivial corrections
to the correlation function in 2D Ising model
vanish in this special case or in the crystallographic
direction $\langle 11 \rangle$, but not in general.

  Our consideration can be generalized easily to the case
where the Hamiltonian parameter $r_0$ is a nonlinear analytical
function of $T$. Nothing is changed in the above expansions
if the reduced temperature $t$, as before, is defined by
$t=r_0(T)-r_0(T_c)$. However, analytical corrections to scaling appear
(and also corrections like\linebreak $(T-T_c)^{m+n \theta}$ with
integer $m$ and $n$) if $t$ is reexpanded in terms of $T-T_c$ at $T>T_c$.
The solution at the critical point remains unchanged, since the phase
transition occurs at the same (critical) value of $r_0$.

\section{Exact transfer matrix algorithms for calculation of the
 correlation function in 2D Ising model}
\label{sec:algorithm}

\subsection{Adoption of standard methods}

The transfer matrix method,
applied to analytical calculations on two--dimensional lattices, is well 
known~\cite{Onsager,Baxter}. The asymptotic behavior of the
correlation functions can be studied by means of
the equations of the conformal field theory~\cite{Henkel}.
Exact equations for the two--point correlation function
of 2D Ising model on an infinite lattice are known, too~\cite{YP}.
However, no analytical methods exist for an exact
calculation of the correlation function in 2D Ising model
on finite--size lattices. 
This can be done numerically by adopting the conventional transfer matrix
method and modifying it to reach the maximal result (calculation of as far 
as possible larger system) with minimal number of arithmetic operations,
as discussed further on.

\begin{figure}
\centerline{\psfig{figure=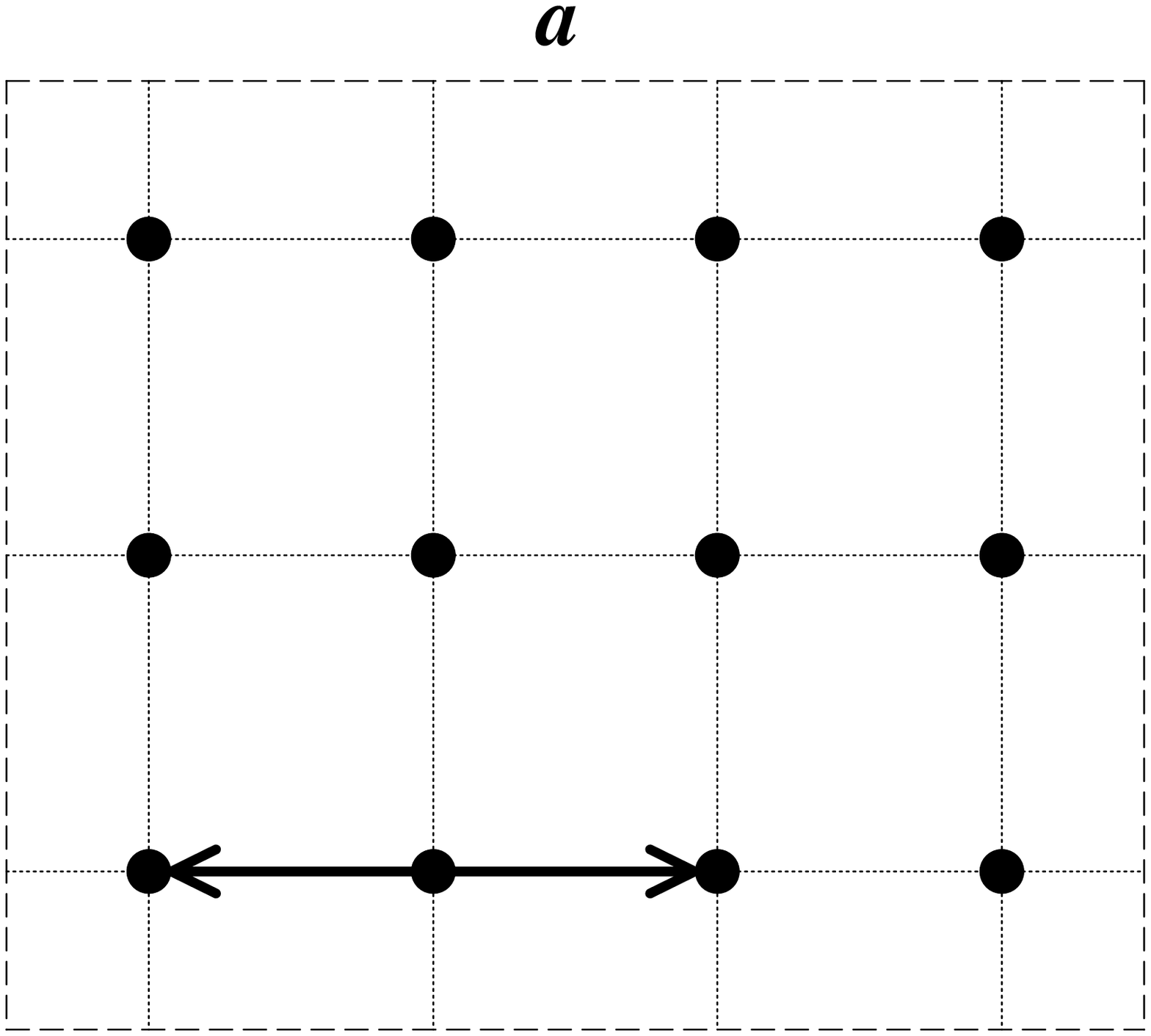,width=6.5cm,height=5.5cm,angle=-90}
	\hspace*{5ex}
            \psfig{figure=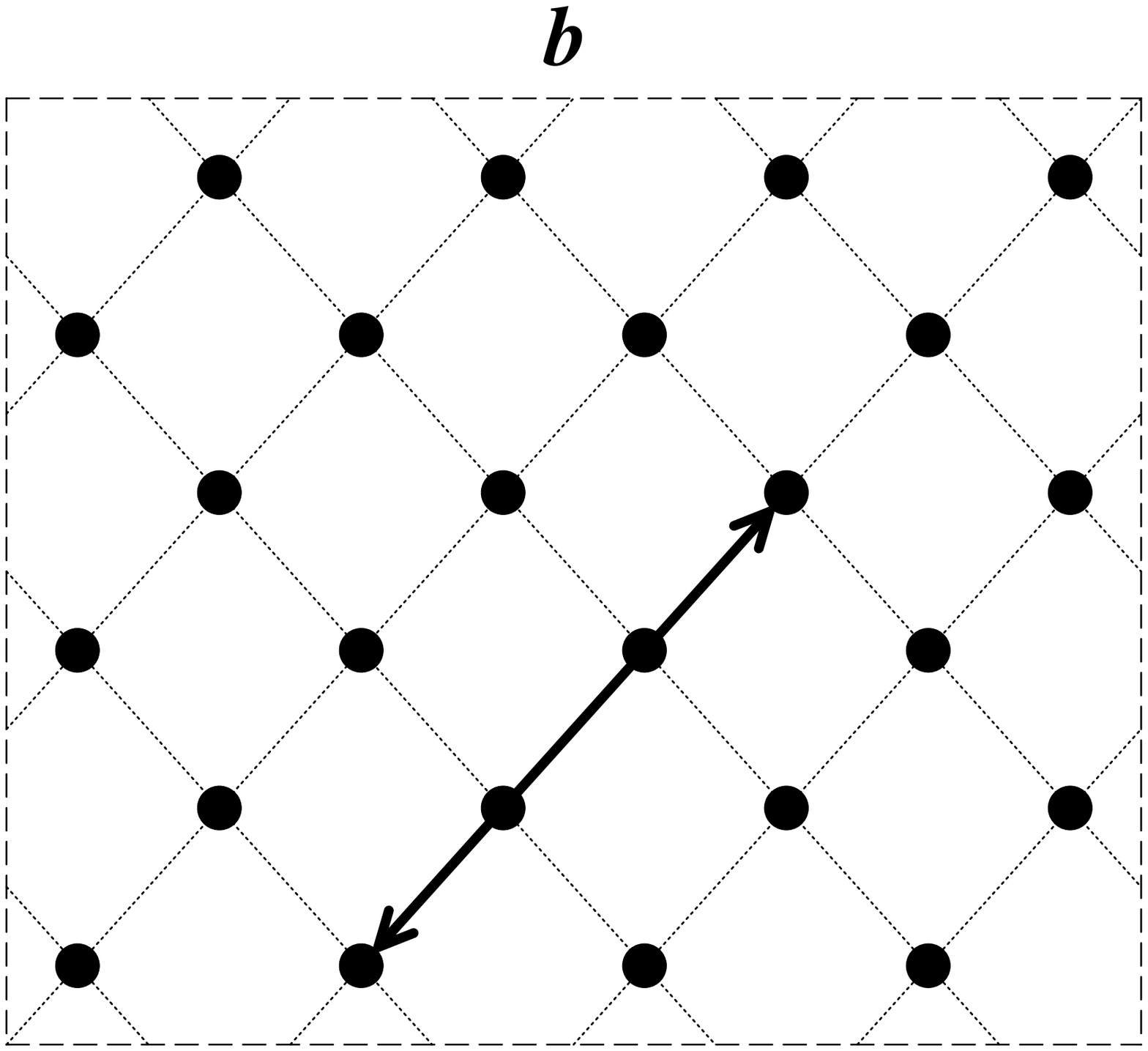,width=6.5cm,height=5.5cm,angle=-90}}
\caption{\small Illustrative examples of the lattices with
dimensions $N \times L$~(a) and $\sqrt{2} N \times \sqrt{2} L$~(b)
with periodic boundary conditions along the dashed lines. 
The correlation function has been calculated in the $\langle 10 \rangle$
crystallographic direction, as indicated by the arrows.}
\label{lattice}
\end{figure}

 We consider the two--dimensional Ising model where spins are located
either on the lattice of dimensions $N \times L$, illustrated in 
Fig.~\ref{lattice}a, or on the lattice of dimensions $\sqrt{2} N \times
\sqrt{2} L$, shown in Fig.~\ref{lattice}b. 
The periodic boundaries
are indicated by dashed lines. In case (a) we have $L$ rows,
and in case (b) -- $2L$ rows, each containing $N$ spins. 
Fig.~\ref{lattice} shows an illustrative example with $N=4$ and $L=3$. 
In our notation nodes are numbered sequently from
left to right, and rows -- from bottom to top.

For convenience, first we consider an application of the transfer matrix
method to calculation of the partition function
\begin{equation}
Z= \sum\limits_{ \{\sigma_k \} } 
\exp \left( \beta \sum\limits_{\langle i,j \rangle} 
\sigma_i \sigma_j \right) \;,
\end{equation}
where $\sigma_i =\pm 1$ are the spin variables, and the summation
runs over all the possible spin
configurations $\{ \sigma_k \}$. The argument of the exponent
represents the Hamiltonian of the system including summation over
all the neighbouring spin pairs $\langle i,j \rangle$ of the given 
configuration $\{ \sigma_k \}$; 
parameter $\beta$ is the coupling constant.
Let us consider lattice~(a) in Fig.~\ref{lattice}, but containing $n$ rows
without periodic boundaries along the vertical axis and without interaction
between spins in the upper row. We define the
$2^N$--component vector ${\bf r}_n$ such that the $i$--th
component of this vector represents the contribution to the partition 
function provided by the $i$--th spin configuration of the upper
row. Then we have a recurrence relation
${\bf r}_{n+1}=T \, {\bf r}_n$,
where $T$ is the transfer matrix which includes the Boltzmann weights
of newly added bonds.
Furthermore, we can write ${\bf r}_{n+1}^{(i)}=T \, {\bf r}_n^{(i)}$,
where ${\bf r}_n^{(i)}$ is the partial contribution to
${\bf r}_n$ provided by the $i$--th configuration of the first row.
The components of ${\bf r}_1^{(i)}$ are given by
$\left( {\bf r}_1^{(i)} \right)_j = \delta_{j,i}$.
In the case of the periodic boundary conditions
the $(L+1)$--th row must be identical to the first one,
which leads to the well known expression~\cite{Baxter,Huang}
\begin{equation}
Z= \sum\limits_i \left( {\bf r}_{L+1}^{(i)} \right)_i = 
\mbox{Trace} \left(T^L \right)= \sum\limits_i \lambda_i^L \;,
\end{equation}
where $\lambda_i$ are the eigenvalues of the transfer matrix $T$.
An analogous expression for the lattice in Fig.~\ref{lattice}b reads
\begin{equation}
Z= \sum\limits_i \left( {\bf r}_{2L+1}^{(i)} \right)_i = \mbox{Trace} 
\left( \left[ T_2 T_1 \right]^L \right) 
%= \sum\limits_i {\lambda'}_i^L 
\;,
\end{equation}
where the vectors ${\bf r}_n^{(i)}$ obey the reccurence relation
${\bf r}_{n+1}^{(i)}=T_{1,2} \, {\bf r}_n^{(i)}$ 
with different transfer matrices $T_1$ and $T_2$ for odd and
even row numbers $n$, respectively.

The actual scheme can be easily adopted to calculate the
correlation function $\langle \sigma_i \sigma_j \rangle$
between any two lattice points $i$ and $j$.
Namely, the correlation funtion $G(x)$ between the points
separated by a distance $x$, like indicated in Fig.~\ref{lattice},
is given by the statistical average $Z' / Z$,
where the sum $Z'$ is calculated in the same way as $Z$, but
including the corresponding product of spin variables. It
implies the following replacements:
\begin{eqnarray}
\left( {\bf r}_1^{(i)} \right)_j = \delta_{j,i} \Rightarrow
\left( {\bf r}_1^{(i)} \right)_j = \delta_{j,i} \,
\left( N^{-1} \sum\limits_{\ell=1}^N \left[ \sigma(\ell) \right]_i
\left[ \sigma(\ell+x) \right]_i \right)&& : \mbox{case~(a)} 
\label{eq:Zpa} \\
\left( {\bf r}_{x+1}^{(i)} \right)_j \Rightarrow 
\left( {\bf r}_{x+1}^{(i)} \right)_j \times
\left( N^{-1} \sum\limits_{\ell=1}^N \left[ \sigma(\ell) \right]_i
\left[ \sigma(\ell+ \Delta(x)) \right]_j \right)&& : \mbox{case~(b)} \;,
\label{eq:Zpb}
\end{eqnarray} 
where 
%the spin variables $\sigma$ refer to the first row,
%and $\sigma'$ -- to the $(1+x)$--th row,
$\Delta(x)=x/2$ holds for even $x$, and $\Delta(x)=(x-1)/2$ --
for odd $x$.
In our notation, $[\sigma(k)]_i$ is the spin variable
in the $k$--th node in a row provided that the whole
set of spin variables of this row forms the $i$--th configuration.
It is supposed that
$\sigma(k+N) \equiv \sigma(k)$ holds according to the periodic
boundary conditions.
Such a symmetrical form, which includes an averaging over $\ell$,
allows to reduce the amount of numerical calculations:
due to the symmetry we need the summation over only $\approx 2^N/N$
nonequivalent configurations of the first row instead of the total
number of $2^N$ configurations.

\subsection{Improved algorithms}

The number of the required arithmetic operations can be further reduced 
if the recurrence relations ${\bf r}_{n+1}^{(i)}=T \, {\bf r}_n^{(i)}$
and ${\bf r}_{n+1}^{(i)}=T_{1,2} \, {\bf r}_n^{(i)}$
are split into $N$ steps of adding single spin. To formulate this
in a suitable way, let us first number all the $2^N$ spin configurations
$\{ \sigma(1); \sigma(2); \cdots ; \sigma(N-1); \sigma(N) \}$ by an
index $i$ as follows:
\begin{equation}
\begin{array}{@{i \: = \:}l@{\hspace{2ex} : \hspace{3ex} \{}*{6}{c}@{\}}}
1 & -1; & -1; & \cdots; & -1; & -1; & -1 \\
2 & -1; & -1; & \cdots; & -1; & -1; & +1 \\
3 & -1; & -1; & \cdots; & -1; & +1; & -1 \\
4 & -1; & -1; & \cdots; & -1; & +1; & +1 \\
\multicolumn{7}{c}\dotfill \\
2^N & +1; & +1; & \cdots; & +1; & +1; & +1 
\end{array}
\label{eq:numbering}
\end{equation}
We remind that the sequence of numbers in the $i$--th row
corresponds to the spin variables
$\left[ \sigma(k) \right]_i$ with $k=1, 2, \ldots, N$.
They change with $i$ just like
the digits of subsequent integer numbers in the binary counting system.

\begin{figure}
\centerline{\psfig{figure=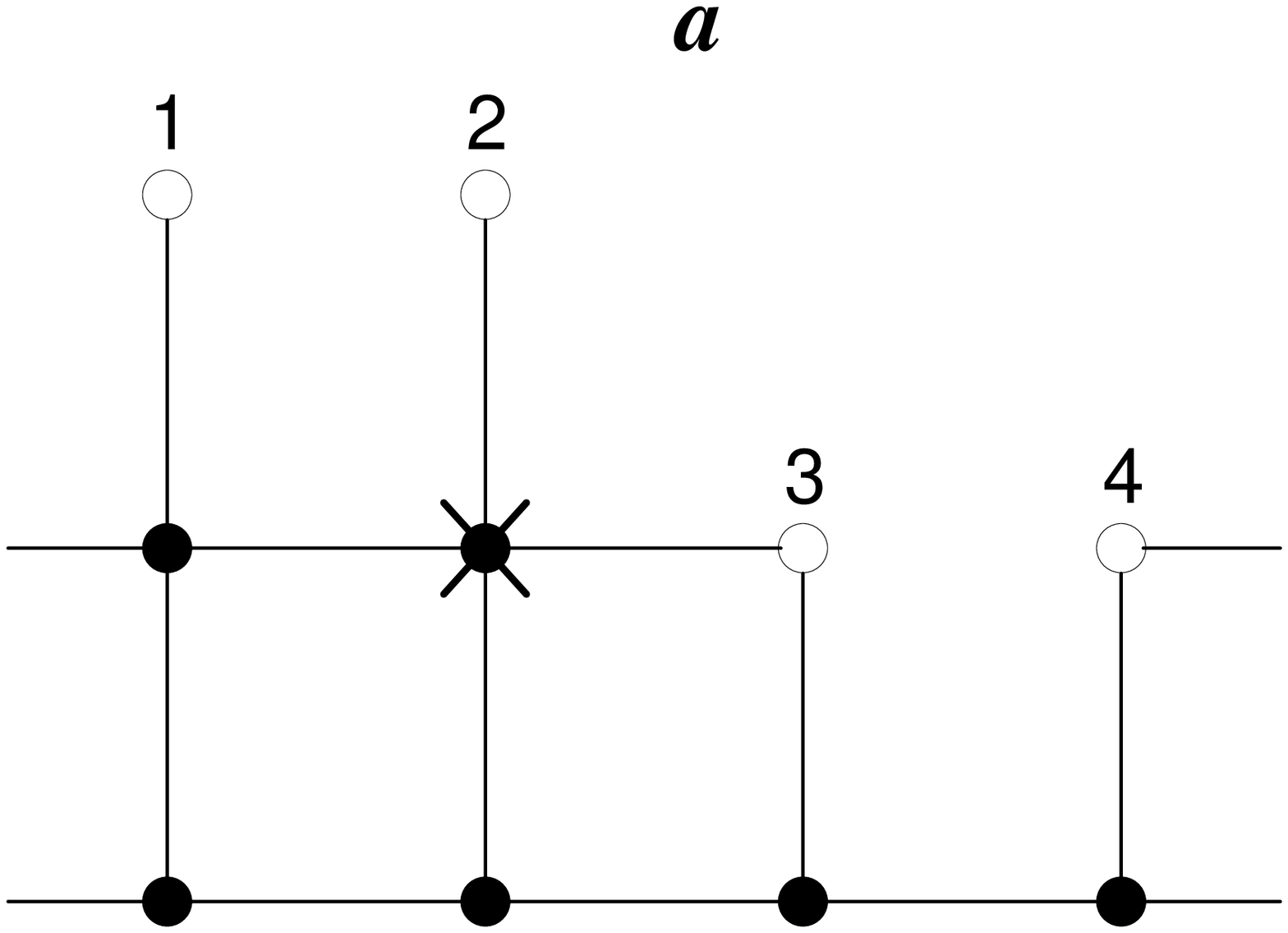,width=6.5cm,height=5.5cm,angle=-90}
	\hspace*{5ex}
            \psfig{figure=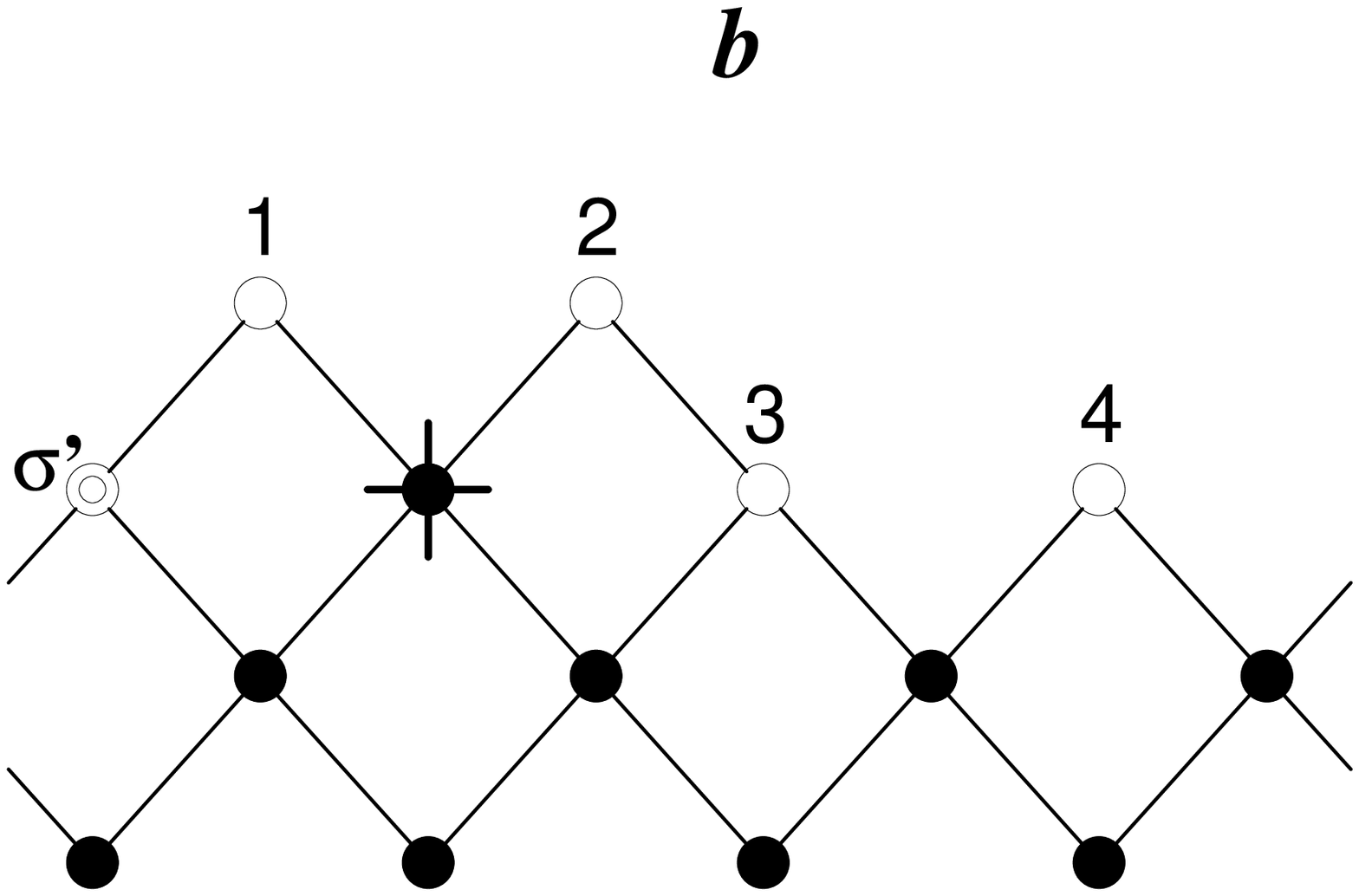,width=6.5cm,height=5.5cm,angle=-90}}
\vspace*{-2ex}
\caption{\small Schematic pictures illustrating the algorithms of 
calculation for the lattices $a$ and $b$ introduced in Fig.~\ref{lattice}.}
\label{latt}
\end{figure}

 Consider now a lattice where $n$ rows are completed, while the
$(n+1)$--th row contains only $\ell$ spins where $\ell < N$, as 
illustrated in Fig.~\ref{latt} in both cases~(a) and~(b) taking
as an example $N=4$. We consider the partial contribution 
$\left( {\bf r}_{n+1,\ell} \right)_i$ (i.~e., $i$--th component of vector
${\bf r}_{n+1,\ell}$) 
in the partition function $Z$ (or $Z'$) provided by a fixed ($i$--th) 
configuration of the set of $N$ upper spins. 
These are the sequently numbered spins
shown in Fig.~\ref{latt} by empty circles.
For simplicity, we have droped 
the index denoting the configuration of the first row. 
In case~(b), the spin depicted
by a double--circle has a fixed value $\sigma'$.
In general, this spin is the nearest bottom--left neighbour of the first 
spin in the upper row. According to this, one has to distinguish 
between odd and even $n$: $\sigma'$ refers either to the first 
(for odd $n$), or to the $N$--th (for even $n$) spin of the $n$--th row. 
It is supposed that the Boltzmann weights are included corresponding
to the solid lines in Fig.~\ref{latt} connecting the spins.
In case~(a) the weights responsible
for the interaction between the upper numbered spins are not included.
Obviously, for a given $\ell>1$, ${\bf r}_{n+1,\ell}$
can be calculeted from ${\bf r}_{n+1,\ell-1}$ via summation over
one spin variable, marked in Fig.~\ref{latt} by a cross.
In case~(a) it is true also for $\ell=1$, whereas in case~(b)
this variable has fixed value $\sigma'$ at $\ell=1$. In the latter case 
the summation over $\sigma'$ is performed at the last step when
the $(n+1)$--th row is already completed. These manipulations enable
us to write 
\begin{equation}
{\bf r}_{n+1}=T \, {\bf r}_n \equiv
\widetilde W_N \, \widetilde W_{N-1} \cdots \widetilde W_2 \,
\widetilde W_1 \, {\bf r}_n 
\label{eq:rec_a}
\end{equation}
with
\begin{equation}
\widetilde W_{\ell} = \sum\limits_{\sigma= \pm 1} W_{\ell} (\sigma) \;,
\label{eq:wa}
\end{equation}
where the componets of the matrices $W_{\ell}(\sigma)$ are given by
\begin{eqnarray}
\left( W_1(\sigma) \right)_{ij} &=& \delta \left( j,j_1(\sigma ,1,i) \right)
\cdot \exp \left( \beta \, \sigma \left\{ \left[ \sigma(1) \right]_i  
+ \left[ \sigma(2) \right]_i + \left[ \sigma(N) \right]_i \right\} \right) 
\nonumber \\
\left( W_{\ell}(\sigma) \right)_{ij} &=& \delta \left( j,j_1(\sigma,\ell,i) 
\right)\cdot \exp \left( \beta \, \sigma \left\{ \left[ \sigma(\ell) \right]_i
+ \left[ \sigma(\ell+1) \right]_i \right\} \right) \hspace{1ex}: 
\hspace{1ex} 1< \ell <N \nonumber \\
\left( W_N(\sigma) \right)_{ij} &=& \delta \left( j,j_1(\sigma,N,i) \right)
\cdot \exp \left( \beta \, \sigma \left[ \sigma(N) \right]_i \right) \;.
\label{eq:a}
\end{eqnarray}
Here $\delta(j,k)$ is the Kronecker symbol and
\begin{equation}
j_1(\sigma,\ell,i)= i+ \left( \sigma- \left[ \sigma(\ell) \right]_i \right)
\, 2^{N-\ell-1}
\label{eq:j1}
\end{equation}
are the indexes of the old configurations containing $\ell-1$ spins in 
the $(n+1)$--th row depending on the value $\sigma$ of the spin
marked in Fig.~\ref{latt}a by a cross, as well as on the index $i$ of the 
new configuration with $\ell$ spins in the $(n+1)$--th row, as consistent
with the numbering~(\ref{eq:numbering}).

The above equations~(\ref{eq:rec_a}) to~(\ref{eq:a}) refers to case~(a). 
In case~(b) we have
\begin{equation}
{\bf r}_{n+1}=T_{1,2} \, {\bf r}_n \equiv \sum\limits_{\sigma'= \pm 1}
\widetilde W_N^{(1,2)} \, \widetilde W_{N-1}^{(1,2)} \cdots 
\widetilde W_2^{(1,2)} \, W_1^{(1,2)}(\sigma') \, {\bf r}_n \;,
\label{eq:rec_b}
\end{equation}
where $\widetilde W_{\ell}^{(1,2)}$ are the matrices 
\begin{equation}
\widetilde W_{\ell}^{(1,2)} = 
\sum\limits_{\sigma= \pm 1} W_{\ell}^{(1,2)} (\sigma) \;.
\label{eq:wb}
\end{equation}
Here indexes 1 and 2 refer to odd and even row numbers $n$, respectively,
and the components of the matrices $W_{\ell}^{(1,2)}(\sigma)$ are 
\begin{equation}
\left( W_{\ell}^{(1,2)}(\sigma) \right)_{ij} 
= \delta \left( j,j_{1,2}(\sigma,\ell,i)
\right) \cdot \exp \left( \beta \left[ \sigma(\ell) \right]_i 
\left\{ \sigma+ \left[ \sigma(\ell+1) \right]_i \right\} \right) \;,
\end{equation}
where $\left[ \sigma(N+1) \right]_i \equiv \sigma'$ and the index 
$j_1(\sigma,\ell,i)$ is given by~(\ref{eq:j1}). For the other index
we have
\begin{eqnarray}
j_2(\sigma,1,i) &=& 2i - 2^{N-1} \left( \left[ \sigma(1) \right]_i +1 \right) 
+ \frac{1}{2} (\sigma-1) \nonumber \\ 
j_2(\sigma,\ell,i) &=& j_1(\sigma,\ell,i) \hspace{2ex}: 
\hspace{2ex} \ell \ge 2 \;.
\end{eqnarray}

Note that the matrices $\widetilde W_{\ell}$ and 
$\widetilde W_{\ell}^{(1,2)}$ have only two nonzero elements in
each row, so that the number of the arithmetic operations required
for the construction of one row of spins via subsequent calculation
of the vectors ${\bf r}_{n+1,\ell}$ increases like 
$2N \cdot 2^N$ instead of $2^{2N}$ operations necassary for a straightforward 
calculation of the vector $T {\bf r}_n$. Taking into account
the above discussed symmetry of the first row, the computation 
time is proportional to $2^{2L}L$ for both $L \times L$ (a)
and $\sqrt{2} L \times \sqrt{2} L$ (b) lattices in Fig.~\ref{lattice}
with periodic boundary conditions.
 
\subsection{Application to different boundary conditions}
\label{subsec:anti}

The developed algorithms can be easily extended
to the lattices with antiperiodic boundary conditions. The latter implies
that $\sigma(N+1)=-\sigma(N)$ holds for each row, and similar condition
is true for each column. We can consider also the mixed boundary
conditions: periodic along the horizontal axis and antiperiodic along
the vertical one, or vice versa. To replace the periodic boundary conditions
with the antiperiodic ones we need only to change the sign of the
corresponding products of the spin variables on the boundaries.
Consider, e.~g., the case~(a) in Fig.~\ref{lattice}. The change of the
boundary conditions along the vertical axis means that the first term
in the argument of the exponent in each of the Eqs.~(\ref{eq:a}) changes 
the sign for the last row, i.~e., when $n=L$. 
The same along the horizontal axis implies
that the term $\left[ \sigma(N) \right]_i$ in the equation for
$\left( W_1(\sigma) \right)_{ij}$ changes the sign. In this case, however,
the symmetry with respect to the configurations of the first row is partly 
broken and, therefore, we need summation over a larger number of  
nonequivalent configurations.

\section{Transfer matrix study of critical Greens function 
and corrections to scaling in 2D Ising model}
\label{sec:correc}

\subsection{General scaling arguments}
\label{subsec:scaling}

It is well known that in the thermodynamic limit the real--space Greens 
function of the Ising model behaves like $G(r) \propto r^{2-d-\eta}$ at 
large distances $r \to \infty$ at the critical point $\beta=\beta_c$, 
where $\eta$ is the critical exponent having the value $\eta=1/4$ in 
two dimensions ($d=2$). Based on our transfer matrix algorithms developed
in Sec~\ref{sec:algorithm}, here we test the finite--size scaling and, 
particularly, the corrections to scaling in 2D Ising model at
the critical point
$\beta= \beta_c= \frac{1}{2} \ln \left( 1+ \sqrt{2} \right)$.

In~\cite{K1} the critical correlation function in 
the Fourier representation, i.~e. $G({\bf k})$ at $T=T_c$, 
has been considered for the $\varphi^4$ model. In this case the minimal 
value of the wave vector magnitude $k$ is related to the linear system size 
$L$ via $k_{min}=2 \pi/ L$. In analogy to the consideration in
Sec.~5.2 of~\cite{K1}, one expects that $k/k_{min}$ is an essential
finite--size scaling argument, corresponding to $r/L$ in the real space.
In the Ising model at $r \sim L$ one has to take into account
also the anisotropy effects, so that the expected finite--size scaling
relation for the real--space Greens function at the critical point
$\beta=\beta_c$ reads
\begin{equation}
G(r) \simeq r^{2-d-\eta} \, f(r/L) \hspace{5ex}:
\hspace{3ex} r \to \infty \;, L \to \infty \;,
\label{eq:scal}
\end{equation}
where the scaling function $f(z)$ depends also on the
crystallographic orientation of the line connecting the correlating spins, 
as well as on the orientation of the periodic boundaries. 
A natural extension of~(\ref{eq:scal}), including the corrections to scaling, 
is
\begin{equation}
G(r) = \sum\limits_{\ell \ge 0} r^{-\lambda_{\ell}} \, f_{\ell}(r/L) \;,
\label{eq:scale}
\end{equation}
where the term with $\lambda_0 \equiv d-2+\eta$ is the leading one,
whereas those with the subsequently increasing exponents $\lambda_1$, 
$\lambda_2$, etc., represent the corrections to scaling.
By a substitution $f_{\ell}(z)=z^{\lambda_{\ell}} f'(z)$, 
the asymptotic expansion~(\ref{eq:scale}) transforms to
\begin{equation}
G(r) = f'_0(r/L) \,  L^{-\lambda_0} \left( 1 + \sum\limits_{\ell \ge 1} 
L^{-\omega_{\ell}} \, \tilde f_{\ell}(r/L) \right) \;,
\label{eq:scalel1}
\end{equation}
where $\tilde f_{\ell}(z) = f_{\ell}'(z)/f_0(z)$ and
$\omega_{\ell}=\lambda_{\ell}-\lambda_0$ are the correction--to--scaling
exponents.

We have tested the scaling relation~(\ref{eq:scal}) in 2D Ising model
by using the exact transfer matrix algorithms in Sec.~\ref{sec:algorithm}.
We have found that all points
of $f(r/L)=r^{1/4}G(r)$ for the correlation function
in $\langle 10 \rangle$ direction (case~(a) in Sec.~\ref{sec:algorithm})
well fit a common smooth line at $2 \le r \le L/2$ and
$L=8$, $12$, $15$, and $18$. It implies that the corrections
to~(\ref{eq:scal}) are rather small.

\subsection{Correction--to--scaling analysis for the
$L \times L$ lattice}
\label{subsec:ll}

Based on the scaling analysis in Sec.~\ref{subsec:scaling},
here we discuss the corrections to scaling for the lattice
in Fig.~\ref{lattice}a. We have calculated the correlation 
function $G(r)$ at a fixed ratio $r/L=0.5$ in $\langle 10 \rangle$ direction, 
as well as at $r/L= 0.5 \sqrt{2}$ in $\langle 11 \rangle$ direction at 
$L=2, 4, 6, \ldots$ with an aim to identify the correction exponents 
in~(\ref{eq:scalel1}). Note that in the latter case the 
replacement~(\ref{eq:Zpb}) is valid for $G(\sqrt{2}x)$ 
(where $x=1, 2, 3, \ldots$) with the only difference that $\Delta(x)=x$. 

Let us define the effective correction--to--scaling exponent 
$\omega_{eff}(L)$ in 2D Ising model via the solution of the equations
\begin{equation}
\tilde L^{1/4}G(r=const \cdot \tilde L) = a + b \, \tilde L^{-\omega_{eff}}
\label{eq:omfit}
\end{equation}
at $\tilde L=L, \, L+\Delta L, \, L+ 2 \Delta L$ with respect to three 
unknown quantities $\omega_{eff}$, $a$, and $b$. 
According to~(\ref{eq:scalel1}), where $\lambda_0=\eta=1/4$,
such a definition gives us the leading correction--to--scaling
exponent $\omega$ at $L \to \infty$, i.~e., $\lim_{L \to \infty} 
\omega_{eff}(L)=\omega$. 

\begin{table}
\caption{\small The critical correlation function $G(r=c \cdot L)$ in $\langle 10 \rangle$ 
($c=0.5$) and $\langle 11 \rangle$ ($c=0.5 \sqrt{2}$) crystallographic 
directions  vs the linear size $L$ of the lattice~(a) in Fig.~\ref{lattice}, 
and the corresponding effective exponents $\omega_{eff}(L)$ and $\widetilde \omega(L)$.}
\label{tab1}
\vspace*{2ex}
\begin{center}
\begin{tabular}{|c|c|c|c|c|c|}
\hline
& \multicolumn{2}{|c|}{\rule[-2.5mm]{0mm}{7mm} direction $\langle 10 \rangle$} 
& \multicolumn{3}{|c|}{direction $\langle 11 \rangle$} \\ \cline{2-6}
\raisebox{1.5ex}{L} 
  & \rule[-3mm]{0mm}{7.5mm}
    $G(0.5 L)$       & $\omega_{eff}(L)$ & $G(0.5 \sqrt{2}L)$ & $\omega_{eff}(L)$ & $\widetilde \omega(L)$ \\ \hline
2 & 0.84852813742386 & 2.7366493         & 0.8                & 1.8672201         &                        \\ 
4 & 0.74052044609665 & 2.9569864         & 0.71375464684015   & 2.2148707         &                        \\
6 & 0.67202206468538 & 1.8998036         & 0.65238484475089   & 2.1252078         &                        \\
8 & 0.62605120856389 & 1.5758895         & 0.60935351016910   & 2.0611362         & 1.909677               \\
10& 0.59238112628953 & 1.6617494         & 0.57724041054810   & 2.0351831         & 1.996735               \\
12& 0.56615525751968 & 1.7774398         & 0.55200680271678   & 2.0232909         & 2.002356               \\
14& 0.54485584658226 & 1.8542943         & 0.53141907668442   & 2.0167606         & 2.001630               \\
16& 0.52703456475995 &                   & 0.51414720882560   &                   &                        \\
18& 0.51178753041103 &                   & 0.49934511003360   &                   &                        \\ \hline
\end{tabular}
\end{center}
\end{table}

The calculated values of $G(r= c \cdot L)$ in the $\langle 10 \rangle$ and 
$\langle 11 \rangle$ crystallographic directions [in case~(a)] with $c=0.5$ 
and $c=0.5 \sqrt{2}$, respectively, and the corresponding
effective exponents $\omega_{eff}(L)$, determined at $\Delta L=2$, are given 
in Tab.~\ref{tab1}. In both cases the effective exponent $\omega_{eff}(L)$
seems to converge to a value about $2$. Besides, in the second case the
behavior is smoother, so that we can try someway to extrapolate
the obtained sequence of $\omega_{eff}$ values (column 5 in Tab.~\ref{tab1})
to $L= \infty$. For this purpose we have considered the ratio of two
subsequent increments in $\omega_{eff}$,
\begin{equation}
r(L)= \frac{\omega_{eff}(L+\Delta L)- \omega_{eff}(L)}
           {\omega_{eff}(L)- \omega_{eff}(L-\Delta L)} \;.
\label{eq:rl}
\end{equation}
A simple analysis shows that $r(L)$ behaves like
\begin{equation}
r(L) = 1 - \Delta L \cdot (\omega'+1) L^{-1} + \mathcal{O} \left( L^{-2} \right)
\label{eq:rapp}
\end{equation}
at $L \to \infty$ if $\omega_{eff}(L) = \omega + \mathcal{O} \left( L^{-\omega'}
\right)$ holds with an exponent $\omega'>1$. 
The numerical data in Tab.~\ref{tab1} show that Eq.~(\ref{eq:rapp})
represents a good approximation for the largest values of $L$ at $\omega'=2$.
It suggests us that the leading and the subleading correction exponents 
in~(\ref{eq:scalel1}) could be $\omega \equiv \omega_1=2$ and $\omega_2=4$, 
respectively. Note that $\omega_{eff}(L)$ can be defined with a shift in
the argument. Our specific choice ensures
the best approximation by~(\ref{eq:rapp}) at the actual finite $L$ values.

Let us now assume that the values of $\omega_{eff}(L)$ are known up to
$L=L_{max}$. Then we can calculate from~(\ref{eq:rl}) the $r(L)$ values
up to $L=L_{max}-\Delta L$ and make a suitable ansatz like
\begin{equation}
r(L)= 1 - 3 \Delta L \cdot L^{-1} + b \, L^{-2} 
\hspace{3ex} \mbox{at} \hspace{2ex} L \ge L_{max}
\label{eq:rapp1}
\end{equation}
for a formal extrapolation of $\omega_{eff}(L)$ to $L=\infty$.
This is consistent with~(\ref{eq:rapp}) where $\omega'=2$. The
coefficient $b$ is found by matching the result to the precisely
calculated value at  $L=L_{max}-\Delta L$. 
The subsequent values of $\omega_{eff}(L)$, calculated
from~(\ref{eq:rl}) and~(\ref{eq:rapp1}) at $L>L_{max}$, converge
to some value $\widetilde \omega(L_{max})$ at $L \to \infty$.
If the leading correction--to--scaling exponent $\omega$
is $2$, indeed, then the extrapolation result
$\widetilde \omega(L_{max})$ will tend to $2$ at $L_{max} \to \infty$
irrespective to the precise value of $\omega'$.

As we see from Tab.~\ref{tab1}, the values of $\widetilde \omega(L)$ 
come remarkably closer to $2$ as compared to $\omega_{eff}(L)$, 
suggesting that $\omega=2$. As we have discussed in
Sec.~\ref{sec:crex}, there could be a nontrivial correction 
in~(\ref{eq:scalel1}) with $\omega= \eta=1/4$.
If it really exists, then it has a very small amplitude,
otherwise it would show up in our analysis.

\subsection{Correction--to--scaling analysis for the
$\sqrt{2} L \times \sqrt{2} L$ lattice}
\label{sec:result}

To test the possible existence of nontrivial corrections to scaling, 
here we make the analysis of the correlation function $G(r)$ in
$\langle 10 \rangle$ direction on the $\sqrt{2} L \times \sqrt{2} L$
lattice shown in Fig.~\ref{lattice}b. 
The advantage of case~(b) in Fig.~\ref{lattice}
as compared to case~(a) is that $\sqrt{2}$ times larger lattice corresponds
to the same number of the spins in one row.
Besides, in this case
we can use not only even, but all lattice sizes to evaluate the
exponent $\omega$ from calculations of $G(r=L)$, which means
that it is reasonable to use the step $\Delta L=1$ to evaluate
$\omega_{eff}$ and $\widetilde \omega(L)$ from Eqs.~(\ref{eq:omfit}),
(\ref{eq:rl}) and~(\ref{eq:rapp1}). The results, are given in Tab.~\ref{tab2}.
\begin{table}
\caption{\small The critical correlation function $G(r=L)$ in
$\langle 10 \rangle$ crystallographic direction and the effective exponents
$\omega_{eff}(L)$ and $\widetilde \omega(L)$ vs the linear size $L$ of 
the lattice~(b) in Fig.~\ref{lattice}.}
\label{tab2}
\vspace*{2ex}
\begin{center}
\begin{tabular}{|c|c|c|c|}
\hline
\rule[-3mm]{0mm}{7.5mm}
L & $G(L)$             & $\omega_{eff}(L)$ & $\widetilde \omega(L)$ \\ \hline
2 & 0.8                &                   &                        \\
3 & 0.7203484812087670 &                   &                        \\
4 & 0.6690636562097066 &                   &                        \\
5 & 0.6321925914229602 &                   &                        \\
6 & 0.6037455936471098 &                   &                        \\
7 & 0.5807668304926868 &                   &                        \\
8 & 0.5616046762441826 & 2.066235298       &                        \\
9 & 0.5452468033693456 & 2.043461090       &                        \\
10& 0.5310294874153481 & 2.030235674       & 1.996772124            \\
11& 0.5184950262041604 & 2.022130104       & 1.999333324            \\  
12& 0.5073151480587211 & 2.016864947       & 1.999941357            \\
13& 0.4972468711401118 & 2.013265826       & 2.000036957            \\
14& 0.4881056192765374 & 2.010701166       & 2.000040498            \\
15& 0.4797481011874659 & 2.008811505       & 2.000044005            \\
16& 0.4720609977942179 & 2.007380630       & 2.000053415            \\
17& 0.4649532511721054 & 2.006272191       & 2.000063984            \\
18& 0.4583506666254706 & 2.005396785       & 2.000073711            \\
19& 0.4521920457268738 &                   &                        \\ 
20& 0.4464263594840965 &                   &                        \\ \hline
\end{tabular}
\end{center}
\end{table}
It is evident from Tab.~\ref{tab2} that the extrapolated values of the
effective correction exponent, i.~e. $\widetilde \omega(L)$, come surprisingly
close to $2$ at certain $L$ values. Besides, the ratio of increments 
$r$~[cf.~Eq.~(\ref{eq:rl})] in this case is well approximated by~(\ref{eq:rapp1}),
as consistent with existence of a correction term in~(\ref{eq:scalel1})
with exponent $4$. On the other hand, we can see from Tab.~\ref{tab2} that 
$\Delta \widetilde \omega(L)= \widetilde \omega(L) - 2$ tends to increase in 
magnitude at $L>13$. 
\begin{figure}
\centerline{\psfig{figure=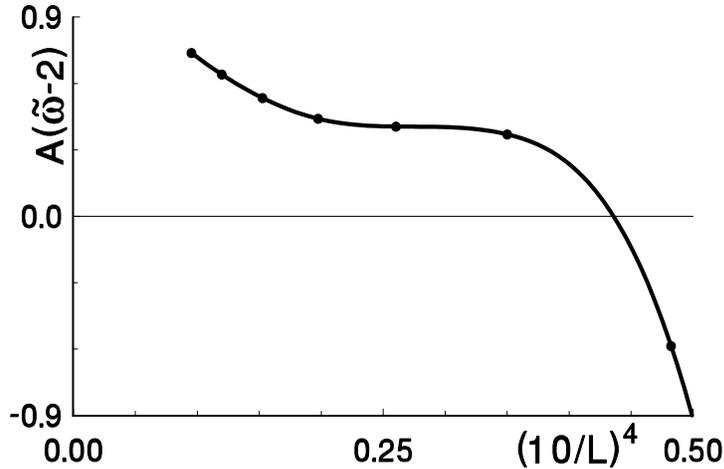,width=11cm,height=8.5cm}}
\vspace{-8ex}
\caption{\small The deviation of the extrapolated effective exponent
$\Delta \widetilde \omega(L)=\widetilde \omega(L)-2$ as a 
function of $L^{-4}$. The extrapolation has been made by using the
calculated $G(r)$ values in Tab.~\ref{tab2} up to the size $L+2$.
A linear convergence to zero would be expected in absence of
any correction term with exponent $\omega<2$.}
\label{fig:dev}
\end{figure}
We have illustrated this systematic and smooth
deviation in Fig.~\ref{fig:dev}. The only reasonable explanation of this behavior
is that the expansion~(\ref{eq:scalel1}) necessarily contains the exponent
$2$ and, likely, also the exponent $4$, and at the same time it contains
also a correction of a very small amplitude with $\omega<2$. The latter
explains the increase of $\Delta \widetilde \omega(L)$. Namely, the
correction to scaling for $L^{1/4}G(L)$ behaves like
$const \cdot L^{-2} \left[ 1 + \mathcal{O} \left( L^{-2} \right) 
+ \varepsilon \, L^{2-\omega} \right]$ with $\varepsilon \ll 1$,
which implies a slow crossover of the effective exponent $\omega_{eff}(L)$
from the values about $2$ to the asymptotic value $\omega$. 
Besides, in the region where $\varepsilon \, L^{2-\omega} \ll 1$ holds,
the effective exponent behaves like
\begin{equation}
\omega_{eff}(L) \simeq 2 + b_1 L^{2-\omega} + b_2 L^{-2} \;,
\label{eq:oeff}
\end{equation} 
where $b_1 \ll 1$ and $b_2$ are constants.
By using the extrapolation of $\omega_{eff}$ with $\omega'=2$ in~(\ref{eq:rapp})
and~(\ref{eq:rapp1}), we have compensated the effect of the correction term 
$b_2 L^{-2}$. Besides, by matching the amplitude $b$ in~(\ref{eq:rapp1}) we
have compensated also the next trivial correction term $\sim L^{-3}$ in
the expansion of $\omega_{eff}(L)$. It means that the extrapolated exponent 
$\widetilde \omega(L)$ does not contain these expansion terms, i.~e., we have
\begin{equation}
\widetilde \omega(L) = 2 + b_1 L^{2-\omega} + \delta \widetilde \omega(L) \;,
\label{eq:omext}
\end{equation} 
where $\delta \widetilde \omega(L)$ represents a remainder term. It 
includes the trivial corrections like $L^{-4}$, $L^{-5}$, etc.,
and also subleading nontrivial corrections,
as well as corrections of order $\left(\varepsilon \, L^{2-\omega} \right)^2$,
$\left(\varepsilon \, L^{2-\omega} \right)^3$, etc.,
neglected in~(\ref{eq:oeff}). According to the latter,
Eq.~(\ref{eq:omext}) is meaningless in the thermodynamic limit 
$L \to \infty$, but it can be used
to evaluate the correction--to--scaling exponent $\omega$ from the
transient behavior at large, but not too large values of $L$ where
$b_1 L^{2-\omega} \ll 1$ holds. In our example the latter condition is well
satisfied, indeed. 

\begin{figure}
\centerline{\psfig{figure=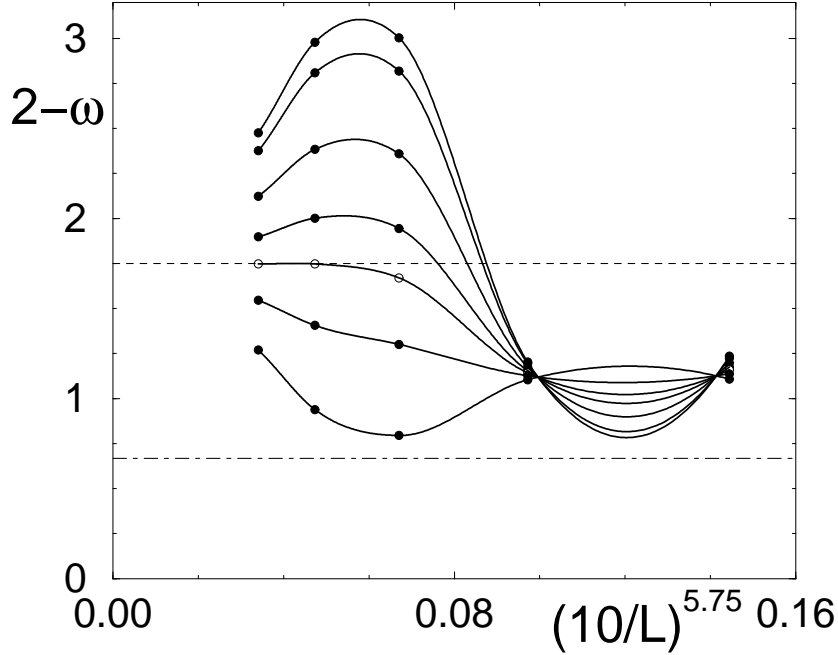,width=11cm,height=8.5cm,angle=-90}}
\caption{\small The exponent $2-\omega$ estimated from~(\ref{eq:lincom})
at different system sizes.
From top to bottom (if looking on the left hand side):
$\alpha = 0, 1, 3.5, 5.75, 7.243, 9.25, 12$. The results at
the optimal $\alpha$ value $7.243$ are shown by empty circles.
The dashed line
indicates our theoretical asymptotic value $2-\omega =1.75$, whereas the
dot--dashed line -- that proposed in~\cite{BF}.}
\label{fig:om}
\end{figure}

Based on~(\ref{eq:omext}), we have estimated
the nontrivial correction--to--scaling exponent $\omega$ by using the
data of $\widetilde \omega(L)$ in Tab~\ref{tab2}.
We have used two different ansatzs
\begin{equation}
2-\omega_1 (L)= \ln \left[ \Delta \widetilde \omega(L) / 
\Delta \widetilde \omega(L-1) \right] / \ln [ L/(L-1)] 
\end{equation}
and
\begin{equation}
2-\omega_2 (L)= L \, \left[ \Delta \widetilde \omega (L) - \Delta \widetilde 
\omega (L-1) \right] / \Delta \widetilde \omega (L) \;,
\end{equation}
as well as the linear combination of them
\begin{equation}
\omega(L)=(1-\alpha) \; \omega_1(L) + \alpha \; \omega_2(L)
\label{eq:lincom}
\end{equation}
containing a free parameter $\alpha$. We have $\omega(L)=\omega_1(L)$
at $\alpha=0$ and $\omega(L)=\omega_2(L)$ at $\alpha=1$. In general,
the effective exponent $\omega(L)$ converges to the same result $\omega$
at arbitrary value of $\alpha$, but at some values the convergence is better.
The results for $2-\omega(L)$ vs $L^{\omega-6}$ at different $\alpha$ values
are represented in Fig.~\ref{fig:om} by a set of curves. 
In this scale the convergence
to the asymptotic value would be linear 
(within the actual region where $L \gg 1$ and
$b_1 \, L^{2-\omega} \ll 1$ hold) for $\alpha=0$ at the condition
$\delta \widetilde \omega(L) \propto L^{-4}$.
We have choosen the scale of $L^{-5.75}$, as it is consistent with our
theoretical prediction in Sec.~\ref{sec:crex} that $\omega=1/4$. Nothing is 
changed essentially if we use slightly different scale as, e.~g., 
$L^{-14/3}$ consistent with the correction--to--scaling exponent $\omega=4/3$ 
proposed in~\cite{BF}.
As we see from Fig.~\ref{fig:om}, all curves tend to merge at our asymptotic
value $2 -\omega=1.75$ shown by a dashed line. The optimal
value of $\alpha$ is defined by the condition that the last two estimates 
$\omega(17)$ and $\omega(18)$ agree with each other. It occurs
at $\alpha=7.243$, and the last two points lie just on our theoretical line.  
A discussion and comparison of our results with those in 
published literature (e.~g.,~\cite{Addendum}) can be found in~\cite{Ktm}.

\subsection{Comparison to the known exact results and estimation
of numerical errors}
\label{sec:comex}

We have carefully checked our algorithms comparing the results
with those obtained via a straightforward counting of all spin
configurations for small lattices, as well as comparing the 
obtained values of the partition function to those calculated
from the known exact analytical expressions. Namely, an
exact expression for the partition function of a finite--size
2D lattice on a torus with arbitrary coupling constants between 
each pair of neighbouring spins has been reported in~\cite{Bednorz} 
obtained by the loop counting method and represented
by determinants of certain transfer matrices. In the standard 2D 
Ising model with only one common coupling constant $\beta$ these 
matrices can be diagonalized easily, using the standard 
techniques~\cite{Landau}. Besides, the loop counting method
can be trivially extended to the cases with antiperiodic
or mixed boundary conditions discussed in 
Sec.~\ref{subsec:anti}. It is necessary only to mention that
each loop gets an additional factor $-1$ when it winds round
the torus with antiperiodic boundary conditions. 
We consider the partition functions
$Z_{pp} \equiv Z$, $Z_{aa}$, $Z_{ap}$, $Z_{pa}$.
In this notation the first index refers to the horizontal
or $x$ axis, and the second one -- to the vertical or
$y$ axis of a lattice illustrated in Fig.~\ref{lattice}a;
$p$ means periodic and $a$ -- antiperiodic boundary conditions.
As explained above, the standard methods leads to the
following exact expressions:
\begin{eqnarray}
Z_{pp} &=& \left( Q_1+Q_2+Q_3-Q_0 \right)/ \, 2 \nonumber \\
Z_{ap} &=& \left( Q_0+Q_1+Q_3-Q_2 \right)/ \, 2 \nonumber \\
Z_{pa} &=& \left( Q_0+Q_1+Q_2-Q_3 \right)/ \, 2 \label{eq:zz} \\
Z_{aa} &=& \left( Q_0+Q_2+Q_3-Q_1 \right)/ \, 2 \nonumber 
\end{eqnarray}
where $Q_0$ is the partition function represented by the sum of 
the closed loops on the lattice, 
as consistent with the loop counting method in~\cite{Landau}, 
whereas  $Q_1$, $Q_2$, and $Q_3$
are modified sums with additional factors
$\exp(\Delta x \cdot i \pi/N + \Delta y \cdot i \pi/L)$, 
$\exp(\Delta x \cdot i \pi/N)$, and 
$\exp(\Delta y \cdot i \pi/L)$, respectively, 
related to each change of coordinate $x$ by $\Delta x = \pm 1$,
or coordinate $y$ by $\Delta y = \pm 1$ when making a loop. 
The standard manipulations~\cite{Landau} yield
\begin{eqnarray} \label{eq:qi}
&&Q_i = 2^{NL} \prod\limits_{q_x, \, q_y} \left[ \cosh^2 (2 \beta)
-\sinh(2 \beta)  \right. \\
&&\left. \times \left( \cos \left[ q_x+ 
\left( \delta_{i,1}+\delta_{i,2} \right) \frac{\pi}{N} \right]
+ \cos \left[ q_y+ \left( \delta_{i,1}+\delta_{i,3} \right) 
\frac{\pi}{L} \right] \right) \right]^{1/2} \nonumber \;,
\end{eqnarray}
where the wave vectors $q_x=(2 \pi/N) \cdot n$ and 
$q_y=(2 \pi/L) \cdot \ell $ run over all the values
corresponding to $n=0, 1, 2, \ldots , N-1$
and  $\ell=0, 1, 2, \ldots , L-1$.
%%%%%%%%%%%%%%%%%%%%%%%%%%%%%%%%%%%%%%%%%%%%%%%%%%%%%%%%%
Eq.~(\ref{eq:qi}) represents an analytic extension
from small $\beta$ region~\cite{Bednorz}. The correct sign
of square roots is defined by this condition, and 
all $Q_i$ are positive except for $Q_0$, which
vanishes at $\beta=\beta_c$ and becomes negative at $\beta>\beta_c$.
%%%%%%%%%%%%%%%%%%%%%%%%%%%%%%%%%%%%%%%%%%%%%%%%%%%%%%%%%%
In the case of the periodic boundary conditions,
each loop of $Q_0$ has the sign $(-1)^{m+ab+a+b}$~\cite{Bednorz},
where $m$ is the number of intersections, $a$ is
the number of windings around the torus in $x$ direction, and 
$b$ -- in $y$ direction. The correct result for $Z_{pp}$ 
is obtained if each of the loops has the sign $(-1)^m$.
In all other cases, similar relations 
are found easily, taking into account the above defined
additional factors. Eqs.~(\ref{eq:zz}) are then obtained
by finding such a linear combination of quantities $Q_i$ which ensures
the correct weight for each kind of loops.

All our tests provided a perfect agreement between the
obtained values of the Greens functions $G(r)$ 
(a comparison between straightforward calculations and
our algorithms), as well as between partition
functions for different boundary conditions
(a comparison between our algorithms and Eq.~(\ref{eq:zz})). 
The relative discrepancies were extremely small (e.~g., $10^{-15}$),
obviously, due to the purely numerical inaccuracy.

We have used the double--precision FORTRAN programs. 
The numerical errors in Tab.~\ref{tab2} have been estimated by repeating
some calculations with twice larger number of
digits (REAL*16 option). Thus, the errors in the $G(L)$ values
for $L=10$ to $L=17$ are $4.7 \cdot 10^{-17}$, $4.06 \cdot 10^{-16}$,
$-3.52 \cdot 10^{-16}$, $-5.65 \cdot 10^{-16}$, $1.03 \cdot 10^{-15}$,
$1.41 \cdot 10^{-15}$, $-1.71 \cdot 10^{-16}$, and $3.09 \cdot 10^{-16}$.
To eliminate the summation error for
the largest lattice $L=20$, we have split the summation over the
configurations of the first row in several
parts in such a way that a relatively small part, 
including only the first 10~000 configurations from the total number
of 52~487 nonequivalent ones, gives the main contribution
to $Z$ and $Z'$.
The same trick with splitting in two approximately
equal parts has been used at $L=19$. As a result, the numerical
errors at $L=18, 19, 20$ are not much larger than the above listed
values for $10 \le L \le 17$. Hence, the resulting
numerical errors in Fig.~\ref{fig:om} do not much exceed $0.03$
in the middle part around $2-\omega \sim 1.75$. 
In Fig.~\ref{fig:dev}, the errors are practically not seen.

\section{Generation of pseudo--random numbers}
\label{sec:random}

We have found that some of simulated quantities like specific heat
of 3D Ising model near criticality are rather sensitive to the quality 
of pseudo--random numbers. The linear congruatial generators
providing the sequence
\begin{equation}
I_{n+1}=(aI_n + c) \mathrm{mod} \, m
\end{equation}
of integer numbers $I_n$ is a convenient choice.
We have used in previous section the generator of~\cite{FMM}
with $a=843314861$, $c=453816693$, and $m=2^{31}$.
The G05CAF generator of NAG library with $a=13^{13}$, $c=0$, and
$m=2^{59}$ (generating odd integers) has been extensively used in~\cite{HasRev}.
We have compared the results of both generators for 3D Ising model, 
simulated by the Wolff's cluster algorithm~\cite{Wolff}, and have 
found a disagreement by almost $1.8\%$ in the maximal value of $C_V$ at the system
size $L=48$. Application of the standard shuffling
scheme~(\cite{MC} p.~391) with the length of the shuffling box (string)
$N=128$ appears to be not helpful to remove the discrepancy.
The problem is that the standard shuffling scheme, where the numbers
created by the original generator are put in the shuffling box and
picked up from it
with random delays in about $N$ steps, effectively removes the
short--range correlations between the pseudo--random numbers, but nevertheless 
it does not essentially
change the block--averages $\langle I_n \rangle_k= k^{-1} \sum_{j=n}^{n+k-1} I_j$
over $k$ subsequent steps if $k \gg N$. It means that
such a shuffling is unable to modify the low--frequency tail of the
Fourier spectrum of the sequence $I_n$ to make it more consistent
with white noise (an ideal case).
The numbers $I_n$ repeat
cyclically and the block--averages over the cycle do not fluctuate
at all in contradiction with truly random behavior.
To solve the problem, we have made a second shuffling as follows.
We have split the whole cycle of length $m$ of the actual generator
with $m=2^{31}$ in $2^{20}$ segments each consisting of 2048 numbers.
Starting with 0, we have recorded the beginning numbers of each segment.
It allows to restart the generator from the beginning of any segment.
The last pseudo--random number generated by our shuffling
scheme is used to choose the next number from the shuffling box,
exactly as in the standard scheme. In addition, we have used the
last but one number to choose at random a new segment after each
2048 steps. This double--shuffling scheme mimics the true fluctuations
of the block--averages even at $k \gg m$ and has an extremely long
(comparable with $\left( 2^9 \atop 2^{20} \right) \left( 2^{20} \right)!$
steps at $N=2^{20}$) cycle.
We have used a very large shuffling box with $N=2^{20}$
to make the shuffling more effective.
As a consequence, we have reached a perfect agreement with the results
of G05CAF generator, which has a rather long cycle even without shuffling.

A hidden problem is the existence of certain long--range correlations
in the sequence $I_n$ of the original generator of~\cite{FMM}.
Namely, pseudo--random numbers of a subset, composed by picking up
each $2^{k}$--th element of the original sequence, appear to be 
rather strongly correlated for $k \ge 20$. It is observed explicitely
by plotting such a subsequence $I^*_n$ vs $n$, particularly, if the first
element is choosen $I^*_1=0$. These correlations reduce the effectiveness
of our second shuffling. Correlations of this kind, although well masked,
exist also in the sequence of G05CAF generator. Namely, if we choose
$I^*_1=1$ and $k=25$ and generate the coordinates ($x, y$) by means
of this subset, then we observe that the $x-y$ plane is filled by 
the generated points in a non--random way. The origin of these correlations, 
obviously, is the choice of modulo parameter $m$ as a power of $2$.
It, evidently, is the reason for systematical errors in some 
applications with Swendsen--Wang algorithm discussed in~\cite{OS}.
A promising alternative, therefore, is to use the well known
Lewis generator~\cite{MC}, where $m=2^{31}-1$ is a prime number, $a=7^5$, 
and $c=0$, as the original generator of our double--shuffling scheme.
(This generator has been tested in~\cite{FLtest} and, 
even without any shuffling, it gave good results for the energy and specific heat
of 2D Ising model on $16 \times 16$ lattice simulated by Wolff's cluster algorithm.) 
As before, the cycle is split in $2^{20}$ segments. However,
the first segment now starts with 1. Besides, 
the first and the last segments contain only 2047 elements instead of 2048. 
After all numbers of the previous segment are exhausted, a new segment is choosen 
as follows: if the last but one random number of our shuffling
scheme is $I$, then we choose the $k$--th segment, where $k=1+ \left[ I/2048 \right]$.
Since we never have $I=0$ or $I=m$, it ensures that each segment is choosen
with the probability proportional to its length. We have used
the shuffling box of length $N=10^6$ for this scheme.

From the theoretical point of view, the latter scheme could provide the
best pseudo--random numbers. The test simulations we made in 2D Ising model
showed that G05CAF generator, as well as both shuffling schemes provide
very accurate results, which indicates that the actually discussed 
long--range correlations do not have a remarkable effect in our 
application.
We have simulated by the Wolff's algorithm the mean energy 
$\langle \varepsilon \rangle$, specific heat $C_V$, as well as its 
derivatives $C'_V=\partial C_V/\partial \beta$ and 
$C''_V=\partial^2 C_V/\partial \beta^2$ for 2D Ising model at the critical point
and have compared the results with those extracted
from exact formulae~(\ref{eq:zz}) and~(\ref{eq:qi}).
The test simulations consisting of $4.8 \cdot 10^8$ and $2.4 \cdot 10^7$ 
cluster--algorithm steps have been made for the lattice sizes $L=48$ and
$L=256$, respectively. The whole simulation has been split in 24 blocks
to estimate the statistical averages and standard errors ($\sigma$) from the last
20 blocks. The simulation with the generator of~\cite{FMM} has revealed
systematical errors of about $10 \sigma$ for the specific heat and its derivatives
at $L=48$. The values provided by
the G05CAF generator and our two shuffling schemes agreed with the exact ones within 
the errors about one $\sigma$. The most serious deviation 
of $2.37 \sigma$ has been observed for $C''_V$ in the case of $L=48$
simulated by our first shuffling scheme. At $L=48$, one standard deviation $\sigma$
corresponded to $\sim 0.0009\%$ relative error for $\langle \varepsilon \rangle$,
$\sim 0.02\%$ error for $C_V$, $\sim 0.2\%$ error for $C'_V$, and $\sim 0.35\%$ error 
for $C''_V$. At $L=256$ these errors were $\sim 0.0012\%$, $\sim 0.12\%$, $\sim 3\%$, 
and $\sim 4\%$, respectively. Furthermore we have used the latter three generators
in simulations of 3D Ising model and verification of the simulated values by performing
some of the simulations twice with different generators.

\section{Test estimations of the critical exponent $\beta$ in 2D
Ising model}
\label{sec:test}

Based on the well known exact magnetization data
$M= \left( 1- [\sinh 2K ]^{-4} \right)^{\beta}$
of 2D Ising model,
we have tested the known method of effective
exponents~\cite{Ballesteros,Janke} extensively used in our paper.
Here $\beta=1/8$ is the magnetization exponent
and $K$ is the coupling constant, denoted in this way
exceptionally in Secs.~\ref{sec:test} to~\ref{sec:mag}.

\begin{figure}
\centerline{\psfig{figure=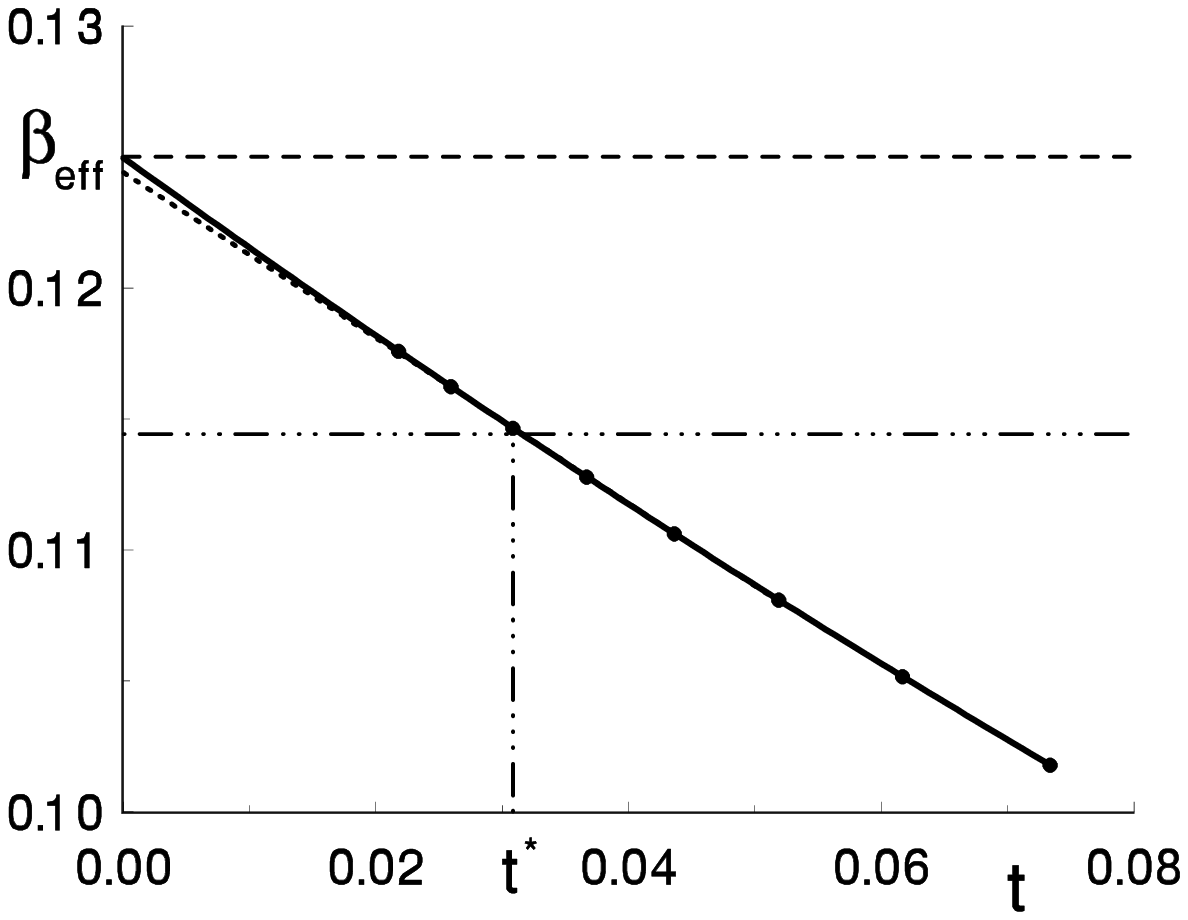,width=8.9cm,height=7.4cm}
        \hspace*{-11ex}
            \psfig{figure=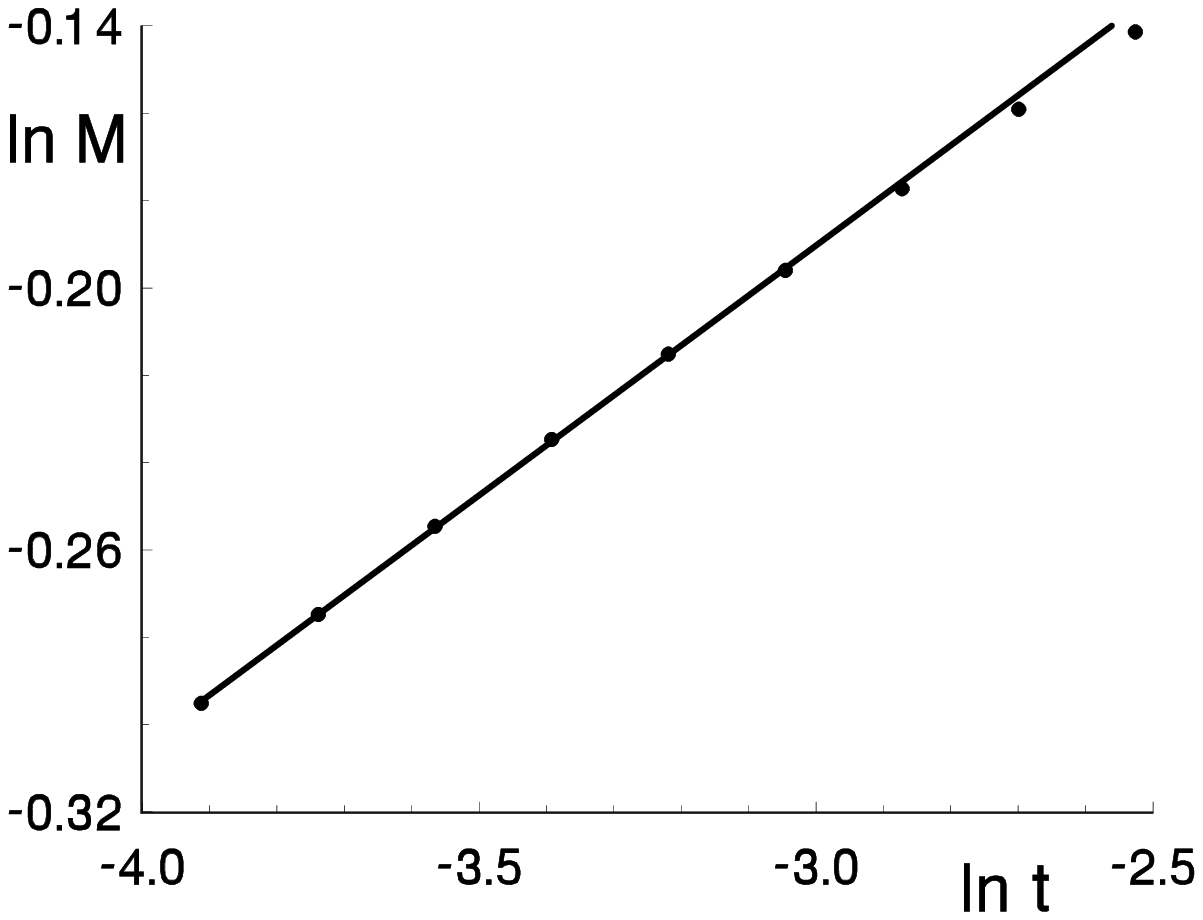,width=8.9cm,height=7.4cm}}
\vspace{-8.5ex}
\caption{\small A test estimation of the critical exponent
$\beta$ in 2D Ising model by the method of effective exponents
(left) and by measuring the slope of the $\ln M$ vs $\ln t$ plot
(right). The tiny--dashed and the solid lines in the left--hand--side
picture show the linear and the quadratic extrapolations
yielding $\beta \simeq 0.1244$ and $\beta \simeq 0.12496$, respectively.
The horizontal dashed line shows the exact value $1/8$.
The slope of the linear $6$--point fit within
$t \in [t_{min};t_{max}]$
in the right--hand--side picture gives 
$\beta \simeq 0.1144 \simeq \beta_{eff}(t^*)$,
where the values $0.1144$ and 
$t^* = \sqrt{t_{min} t_{max}}$ are indicated in the
other picture by dot--dot--dashed lines.}
\label{beta2d}
\end{figure}
The effective critical exponent
$\beta_{eff}(t) = \ln[ M(at) / M(t/a)] / (2 \ln a)$
with $a= 2^{1/4}$, calculated from the magnetization data
$M(t)$ within the range of the reduced temperatures
$t = (K/K_c)-1 \in [0.02;0.08]$ (where $K_c$
is the critical coupling) is shown in
Fig.~\ref{beta2d} (left) by solid circles.
Omitting two largest $t$ values, the linear least--squares
approximation of $\beta_{eff}(t)$ (tiny dashed line)
gives $\beta \simeq 0.1244$, and the quadratic
fit of all points (solid line) yields $\beta \simeq 0.12496$
in close agreement with the exact value $0.125$.
For comparison, the most popular method of estimation
of critical exponents by simply measuring the slope
of a log--log plot, as illustrated in the 
right--hand--side picture (Fig.~\ref{beta2d}),
yields a relatively poor result
$\beta \simeq 0.1144$ in spite of the fact that the actually
used piece of the log--log plot (6 smallest $t$ values)
looks very linear. Up to now we have discussed the exact data only.
In the case of Monte Carlo data with the statistical errors, say,
about the symbol size, we would be unable to detect
the very small deviations from linearity and could
easily get a very good, but nevertheless misleading linear fit. 
In other words, such a simple measuring of critical
exponent is unreliable since there is clearly a danger
to get an uncontrolled systematical
error. Such a measurement within
$t \in [t_{min};t_{max}]$ practically yields the
mean slope of the log--log plot within this interval, which
is nothing but an effective exponent. 
It corresponds just to one point on
the $\beta_{eff}(t)$ plot, i.~e., to
$\beta_{eff}(t^*)$ at $t^* = \sqrt{t_{min} t_{max}}$, as
indicated in Fig.~\ref{beta2d} (left) by dot--dot--dashed lines.
The method of effective exponents has been designed to control
the systematical errors of such simple measurements and eliminate
them by a suitable extrapolation.
Inclusion of corrections to scaling directly in the ansatz for
row data (in this case $M$) is another way to eliminate the
systematical errors.
However, we greatly prefer the method of effective exponents
since it can be well controlled visually.
This method is also sensitive enough to distinguish between
a power--like and a logarithmic singularity of specific heat,
as discussed in Sec.~\ref{sec:alfa}.

There are no essential problems reported in literature, as regards
the MC estimation of the critical exponent $\beta$ in 2D Ising model.
It is because the simulations can be done easily much closer
to the critical point than in our test example.
However, the problem remains in 3D case. The systematical
errors in the measured $\beta$ values are caused by the corrections
to scaling, so that $t^{\theta}$ rather than $t$ is
an essential parameter.
For the smallest reduced temperatures $t \ge 0.0005$
considered in the published literature~\cite{TB} we have
$t^{\theta} > 0.022$ with the RG value
of the correction--to--scaling exponent $\theta \simeq 0.5$,
and $t^{\theta} > 0.079$ with our (GFD) value $\theta=1/3$.
It means that the systematical errors of
the simple (naive) measurements of $\beta$ can be even larger
than in our 2D test example with $t^{\theta} \equiv t \ge 0.02$.

%\newpage

\section{Estimation of the critical coupling of 3D Ising model}
\label{sec:critp}

Here we discuss the critical coupling $K_c$ of 3D Ising model,
which is relevant to our estimations of the critical exponent
$\beta$ in Sec.~\ref{sec:mag}.
The most accurate MC values reported in literature are
$K_c=0.22165459(10)$~\cite{BST99}, $K_c \simeq 0.2216545$~\cite{HV},
and $K_c = 0.2216544(3)$~\cite{TB}. One of the recent
estimates of HT series expansion is $K_c=0.221654(1)$~\cite{BC2}.
We have estimated the critical coupling from our MC results for
the pseudocritical couplings $\tilde K_c(L)$ which correspond to
$U=1.6$, where $U= \langle M^4 \rangle / \langle M^2 \rangle^2$.
Note that $1-U/3$ is the Binder cumulant which may have the values
from $0$ (at high temperatures) to $2/3$ (at low temperatures).
In the thermodynamic limit $L \to \infty$ it changes jump--likely
at $K=K_c$, so that $\lim\limits_{L \to \infty} \tilde K_c(L) = K_c$
holds for any given $U$ within $1< U < 3$.

The values
$\tilde K_c(48)=0.22164540(118)$,
$\tilde K_c(64)=0.22165095(153)$,
$\tilde K_c(96)=0.221653069(734)$,
$\tilde K_c(128)=0.221653945(453)$,
$\tilde K_c(192)=0.221654550(316)$,
\linebreak $\tilde K_c(256)=0.221654755(163)$, and
$\tilde K_c(384)=0.221654532(109)$
have been obtained by an iterative method similar to that
described in Sec.~\ref{sec:alfa}.
%The simulations are
%still continued and the details, including refined values,
%will be given elsewhere.

The data suggest that the pseudocritical coupling has a maximum
at $L \sim 256$. Since $dU/dK$ is negative, such a qualitative
behavior is expected in view of the known results~\cite{Has},
according to which the universal value of $U$ at $K=K_c$ and
$L \to \infty$ is slightly larger than $1.6$ and, therefore,
$\tilde K_c(L)$ should approach $K_c$ from above.
According to the finite--size scaling theory,
$\tilde K_c(L) - K_c \propto L^{-1/\nu} \left[ 1+ \mathcal{O}
\left( L^{-\omega} \right) \right]$ holds at large $L$,
where $\omega=\theta/\nu$.
We have found that the data within a wide range of sizes
can be well described by the Pade approximation
rather than by a simple ansatz with the correction--to--scaling term.
Namely, the formula
\begin{equation}
\tilde K_c(L) \simeq K_c + \mathcal{L}^{-1/\nu} \;
\frac{a_0 + a_1 \mathcal{L}^{-\omega}}{1+ \mathcal{L}^{-\omega}} \;,
\label{eq:Pade}
\end{equation}
where $\mathcal{L} = L/L_0$ and $L_0$ corresponds to the
maximum of $\tilde K_c(L)$ plot, well fits the data within
the whole range of sizes $L \in [48;384]$.
The location of the maximum is the only characteristic length
measure for the $\tilde K_c(L)$ plot, which should transform
to the correct asymptotic form somewhat beyond this maximum.
It motivates our specific choice of $L_0$, which otherwise
is not well defined as a fitting parameter.
Fortunately, the results remain practically
unchanged if we take, say, twice smaller or twice larger value
of $L_0$.

Assuming our (GFD) exponents $\nu=2/3$ and 
$\omega=0.5$ (correction--to--scaling exponent for the magnetization),
a fitting of all data points to~(\ref{eq:Pade})
yields $K_c = 0.22165407(29)$ with the goodness of fit~\cite{NumRec}
$Q=0.897$. To eliminate the systematical errors,
we have discarded the two smallest sizes, which yields
$K_c=0.22165386(51)$ with $Q=0.797$.
It agrees within the error bars  with the value
$0.2216544(3)$ of~\cite{TB} and the value $0.221654(1)$ of~\cite{BC2}.
Our value is provided by the fit
within $L \in [96;384]$ at $L_0=234$. It is shifted up (down)
only by $5.5 \cdot 10^{-8}$ ($6.1 \cdot 10^{-8}$)
at a twice smaller (larger) $L_0$. It is also not very sensitive
to the choice of the exponents $\nu$ and $\omega$.
Assuming the RG exponents $\nu=0.63$ and $\omega=0.8$, we obtain
$K_c=0.22165395(46)$ with $Q=0.795$.
Further we have used both our values of $K_c$ and those reported
in literature in various estimations of the critical exponent $\beta$.

\section{Estimation of the critical exponent $\beta$ from
the magnetization data in 3D Ising model} \label{sec:mag}

 Based on the well known scaling relation
$2 \beta = d \nu - \gamma$, we find from~(\ref{eq:expo1})
and~(\ref{eq:expo2}), where $j=0$ and $m=3$ holds at $n=1$,
the GFD value $\beta=3/8$ of the magnetization exponent $\beta$ for
3D Ising model.
This value is remarkably larger than the usually believed ones
about $0.326$~\cite{Justin}. We suppose that the asymptotic
exponent $\beta$ not only for the Ising model, but also for the
Heisenberg model is larger than provided by approximate RG
theories and known numerical estimates. In polycrystalline Ni ($n=3$), the
increase of the effective exponent $\beta_{eff}$ well above the RG
value 0.3662~\cite{Justin1} has been established experimentally
in~\cite{SRS},
where the authors have found the asymptotic estimate $\beta=0.390(3)$.
This value clearly disagree within error bars with the RG prediction,
but agree with our value $11/28=0.3928 \ldots$ predicted for
the $n=3$ case ($m=3$, $j=2$). Also the critical exponent $\gamma$
measured in most of experiments on Ni and Fe ranges from
1.28 to 1.35 (see~\cite{SRS,BSC,SMB} and references therein),
and only some experimentators have obtained a larger value
about 1.41 -- the only value cited in~\cite{Justin}.
One of the best experimental methods is
to use the Kouvel--Fisher plot, since $T_c$ and $\gamma$ are
determined simultaneously with no fitting parameters~\cite{BSC}.
This method yields the value $\gamma=1.35$~\cite{BSC,KF}
which is believed among (some) experimentators to be the asymptotic
exponent (see the references in~\cite{BSC}).
Our prediction $\gamma=19/14 \simeq 1.357$ is remarkably consistent
with this value.

 Let us now return to the Ising model.
The spontaneous magnetization $M$ of 3D Ising model 
has been considered in~\cite{IS1,Ito,CH,TB}.
An empirical formula
\begin{equation} \label{eq:TBf}
M(\hat t) = \hat t^{0.32694109} \left( 1.6919045
- 0.34357731 \, \hat t^{0.50842026} -0.42572366 \, \hat t \right)
\end{equation}
with $\hat t = 1-0.2216544/K$
has been found in~\cite{TB} which fits the simulated at
three linear sizes $L=64, 128, 256$ of the lattice
and extrapolated to the thermodynamic limit data of
$\langle \mid M \mid \rangle$ within the range of the reduced
temperatures $\hat t \simeq t=(K/K_c)-1 > 0.0005$.
We have made an approximate estimatiom of the spontaneos
magnetization $M(t)$ in the thermodynamic limit from our simulated
values of $\langle \mid M \mid \rangle$ at $L=200$
to compare the results with~(\ref{eq:TBf}).
%It is possible at not too small values of $t$.
Besides, we have made accurate MC simulations by the Wolff's
algorithm~\cite{Wolff} at $t \ge 0.000086$ for system sizes up
to $L=410$ to verify the critical exponent $\beta \simeq 0.3269$
proposed by~(\ref{eq:TBf}).

We have performed the simulations at certain values of coupling
constants $K_i$,
\begin{equation} \label{eq:grid}
K_i=K_c^* + \frac{0.224-K_c^*}{(\sqrt{2})^i} \;,
\hspace{3ex} -2 \le i \le 14 
\end{equation}
(rounded to 9 digits after the decimal point),
where $K_c^*=0.2216545$ is the critical coupling estimated in~\cite{HV}.
We have choosen $K_0=0.224$ to compare
our results with those reported in~\cite{Ito,CH}. Each next $K_i$ value
is $\sqrt{2}$ times closer to $K_c^*$ than the previous one.

Our MC simulations have been split typically in $51$ blocks (bins) 
to calculate the average value an standard
deviation $\sigma$ from the last $50$ bins.  
As a test, we have checked that
a splitting in twice larger blocks provides consistent values of $\sigma$,
i.~e., the blocks were large enough to justify our treatment of the
block--averages as statistically independent quantities. 
In most of the cases one bin included $J=120000$ cluster updates,
which corresponds to $J \langle M^2 \rangle$  complete updates of the whole 
system or sweeps, as consistent with the known improved cluster
estimator $\langle M^2 \rangle=L^{-d} \langle c \rangle$~\cite{MC},
where $\langle c \rangle$ is the average cluster size.
Somewhat shorter
%(as consistent with the reported statistical errors)
simulations have been performed for our estimations of
$\langle \mid M \mid \rangle$ in Tab.~\ref{magmod}.
The results at our smallest $K$ value $K_{14} = 0.221672824$
have been obtained by an averaging over four runs, i.~e.,
$4 \times 50 \times 60000$ cluster updates.
Note that the auto--correlation time of the Wolff's
algorithm at criticality is only few (2 or 3) sweeps~\cite{Wolff}.
The MC measurements were made frequently with respect to this 
auto--correlation time, i.~e., the fraction of moved spins between 
the measurements was about $0.15$ or smaller.
In all cases we have discarded no less than $300$ sweeps from the
beginning of the simulation.
We have verified that the system has been equilibrated well enough
by comparing the estimates from separate smaller parts of the whole
simulation. 

Our $\langle \mid M \mid \rangle$ data for $K_{-2} \le K \le K_3$
are listed in Tab.~\ref{magmod}. The second shuffling scheme described in
Sec.~\ref{sec:random} has been used as a source of pseudo--random numbers
for this simulation.
\begin{table}
\caption{\small The simulated values of $\langle \mid M \mid \rangle$ 
at four different system sizes $L$ in the range of 
coupling constants $K_3 \le K \le K_{-2}$.}
\label{magmod}
\vspace*{2ex}
\begin{center}
\begin{tabular}{|l|l|l|l|l|}
\hline
\rule[-2mm]{0mm}{7mm}
           & $L=60$         & $L=80$         & $L=120$        & $L=200$       \\ \hline
$K=K_{-2}$ & 0.460754(205)  & 0.460695(149)  & 0.460787(155)  & 0.460493(96)  \\
$K=K_{-1}$ & 0.414874(203)  & 0.414845(236)  & 0.414329(162)  & 0.414503(141) \\
$K=K_0$    & 0.372813(339)  & 0.372785(240)  & 0.372939(212)  & 0.372389(142) \\
$K=K_1$    & 0.334641(378)  & 0.334207(221)  & 0.334246(211)  & 0.334327(122) \\
$K=K_2$    & 0.299917(295)  & 0.299268(298)  & 0.299473(211)  & 0.299621(163) \\
$K=K_3$    & 0.268433(402)  & 0.268930(294)  & 0.268807(286)  & 0.268723(200) \\  \hline
\end{tabular}
\end{center}
\end{table}
The values of $\langle \mid M \mid \rangle$ are only slightly varied with $L$,
and those at the largest two sizes $L=120$ and $L=200$ agree or almost agree
within the error bars. According to~\cite{CH}, the latter size
more than 22 times exceeds the correlation length $\xi$
at $K \ge K_3$, which is quite enough to estimate
the thermodynamic limit.
Our data at $L=200$ are in a reasonable agreement with the corresponding
values 
0.460435, 0.414490, 0.372471, 0.334258, 0.299652, and 0.268412
given by~(\ref{eq:TBf}) at $K=K_{-2}$ to $K=K_3$.
Our result at $K=K_0=0.224$, i.~e. $M =0.372389(142)$ or $M^2=0.138674(106)$,
agree within the error bars with the $M^2$ value $0.138708(39)$ reported
in~\cite{Ito}, as well as with $M^2$ value $0.1387488(75)$ obtained
in~\cite{CH}.

 Our estimation of the critical exponent $\beta$ is based on the
analysis of the effective exponent
\begin{equation} \label{eq:findif}
\beta_{eff}(t) = \frac{\ln \langle M^2(t_1,L(t_1)) \rangle
- \ln \langle M^2(t_2,L(t_2)) \rangle }{2(\ln t_1 - \ln t_2)} \;,
\end{equation}
where $\ln t = (\ln t_1 + \ln t_2)/2$, $L(t)t^{\nu}= const$,
and $\langle M^2(t,L) \rangle$ is the statistically
averaged squared magnetization at the given $t$ and $L$.
The effective exponent is the average slope of the
$0.5 \ln \langle M^2 \rangle$ vs
$\ln t$ plot within $t \in [t_1;t_2]$, calculated at a fixed
scaling argument $L t^{\nu}$ which corresponds to a certain
asymptotic value of  $L/\xi$ at $t \to 0$,
where $\xi \sim t^{-\nu}$ is the correlation length.
For any given ratio $t_2/t_1$, the values of $\beta_{eff}(t)$ lie on 
a smooth analytical curve converging to the true asymptotic exponent
$\beta$ at $t \to 0$. Besides, the estimates obtained at slightly different 
values of $t_2/t_1$ well coincide with each other.
We have choosen $L(t)=256 (\tilde t/t)^{\nu}$, where $\tilde t$
is a reference value of the reduced temperature at
$K=K_{12}=0.221691148$. In this case $L(t)$ approximately corresponds
to the minimum of $\langle \mid M \mid \rangle$ vs $L$ plot~\cite{Private},
and the deviations from the thermodynamic limit are small.
Two slightly different values for the
correlation length exponent $\nu$ have been used, i.~e., $\nu=2/3$ (our value) and
$\nu = 0.63$ (RG value).

We have made the simulations at
$L$ values close to $L(t)$, which allow us to estimate 
$\langle M^2(t,L(t)) \rangle$ both
at $\nu=2/3$ and $\nu=0.63$ by a linear interpolation of $\langle M^2 \rangle$ vs
$L^{-d}$ plot. This plot is linear at $L \to \infty$,
as discussed in~\cite{IS1}. At $K = K_{13}, K_{14}$, 
all (three) points have been fit together to get a more reliable result.
The simulated values are listed in Tab.~\ref{tabs}.
\begin{table}
\caption{\small Squared magnetization $\langle M^2(K,L) \rangle$ at different
coupling constants $K$ and system sizes $L$. The simulated values
have been obtained by using G05CAF pseudo--random number generator, except
those marked by an asterisk, for which our second shuffling scheme 
(Sec.~\ref{sec:random}) has been applied. The value of~\cite{Private}
is marked by a dagger.}
\label{tabs}
\vspace*{2ex}
\begin{center}
\begin{tabular}{|c|l||c|l||c|l|}
\hline
\rule[-2.5mm]{0mm}{7mm}
L  & $\langle M^2(K_{-1},L) \rangle$ & L & $\langle M^2(K_0,L) \rangle $ 
& L & $\langle M^2(K_1,L) \rangle $  \\ \hline
12 & 0.186509(149)   & 16  & 0.148333(120)     & 20 & 0.119325(134) \\
13 & 0.182897(143)   & 18  & 0.144657(144)     & 21 & 0.117712(98)  \\
14 & 0.180291(156)   & 19  & 0.143549(125)     & 23 & 0.116133(116) \\
15 & 0.177886(154)   &     &                   & 24 & 0.115459(127) \\ \hline
%\end{tabular}
%\end{center}
%\begin{center}
%\begin{tabular}{|c|l||c|l||c|l|}
\hline
\rule[-2.5mm]{0mm}{7mm}
L  & $\langle M^2(K_2,L) \rangle$ & L & $\langle M^2(K_3,L) \rangle $ 
& L & $\langle M^2(K_4,L) \rangle $  \\ \hline
25 & 0.095551(72)   & 32  & 0.076381(92)      & 40 & 0.060932(105) \\
26 & 0.095077(113)  & 32  & $0.076290(109)^*$ & 41 & 0.060583(87)  \\
28 & 0.093522(99)   & 35  & 0.074865(99)      & 44 & 0.059919(68)  \\
29 & 0.092933(107)  & 36  & 0.074543(103)     & 45 & 0.059726(92)  \\ \hline
%\end{tabular}
%\end{center}
%\begin{center}
%\begin{tabular}{|c|l||c|l||c|l|}
\hline
\rule[-2.5mm]{0mm}{7mm}
L  & $\langle M^2(K_5,L) \rangle$ & L & $\langle M^2(K_6,L) \rangle $      
& L & $\langle M^2(K_7,L) \rangle$  \\ \hline
50 & 0.048762(79)   & 64  & 0.038874(80)      & 80 & 0.031042(79)  \\
51 & 0.048476(89)   & 64  & $0.038872(69)^*$  & 83 & 0.030865(68)  \\
55 & 0.047911(79)   & 69  & 0.038338(83)      & 86 & 0.030618(69)  \\
56 & 0.047747(75)   & 70  & 0.038252(79)      &    &               \\ \hline
%\end{tabular}
%\end{center}
%\begin{center}
%\begin{tabular}{|c|l||c|l||c|l|}
\hline
\rule[-2.5mm]{0mm}{7mm}
L   & $\langle M^2(K_8,L) \rangle$ & L & $\langle M^2(K_9,L) \rangle $ 
& L & $\langle M^2(K_{10},L) \rangle$  \\ \hline
101 & 0.024673(66)   & 128  & 0.019726(73)      & 161 & 0.015670(49)     \\
104 & 0.024538(60)   & 128  & $0.019735(50)^*$  & 166 & 0.015575(46)     \\
107 & 0.024466(63)   & 133  & 0.019596(59)      &     &                  \\ \hline
%\end{tabular}
%\end{center}
%\begin{center}
%\begin{tabular}{|c|l||c|l||c|l|}
\hline
\rule[-2.5mm]{0mm}{7mm}
L   & $\langle M^2(K_{11},L) \rangle$ & L & $\langle M^2(K_{12},L) \rangle$  
& L & $\langle M^2(K_{13},L) \rangle$  \\ \hline
203 & 0.012498(45)    & 256 & 0.009936(47)             & 315 & 0.007950(32) \\
206 & 0.012464(62)    & 256 & $0.009862(49)^{\dagger}$ & 320 & 0.007911(43) \\
    &                 & 256 & $0.009976(87)^*$         & 320 & $0.007943(39)^*$ \\
    &                 &     &                          & 325 & 0.007901(38) \\ \hline
\hline
\rule[-2.5mm]{0mm}{7mm}
L   & $\langle M^2(K_{14},L) \rangle$ & L & $\langle M^2(K_{14},L) \rangle$  
& L & $\langle M^2(K_{14},L) \rangle$  \\ \hline
390 & 0.006344(25)    & 400 & 0.006317(27) & 410 & 0.006275(25) \\ \hline
\end{tabular}
\end{center}
\end{table}
In most of the cases the G05CAF pseudo--random number generator has been used,
whereas some of the values, which are marked by an asterisk, have been simulated
by our second shuffling scheme (Sec.~\ref{sec:random}).
The first value at $K=K_{12}$ and $L=256$ has been obtained from
$50 \times 90000$ cluster updates. The second one represents the
result of~\cite{Private} extracted from the simulation of
approximately the same length with the same G05CAF generator.
As we see, different simulations confirm each other within the error bars. 
We have made a weighted averaging
(with the weights proportional to the simulation length which is
roughly $\propto 1/\sigma^2$)
of the overlaping simulation results to obtain statistically 
more reliable values for our further analysis.

\begin{figure}
\centerline{\psfig{figure=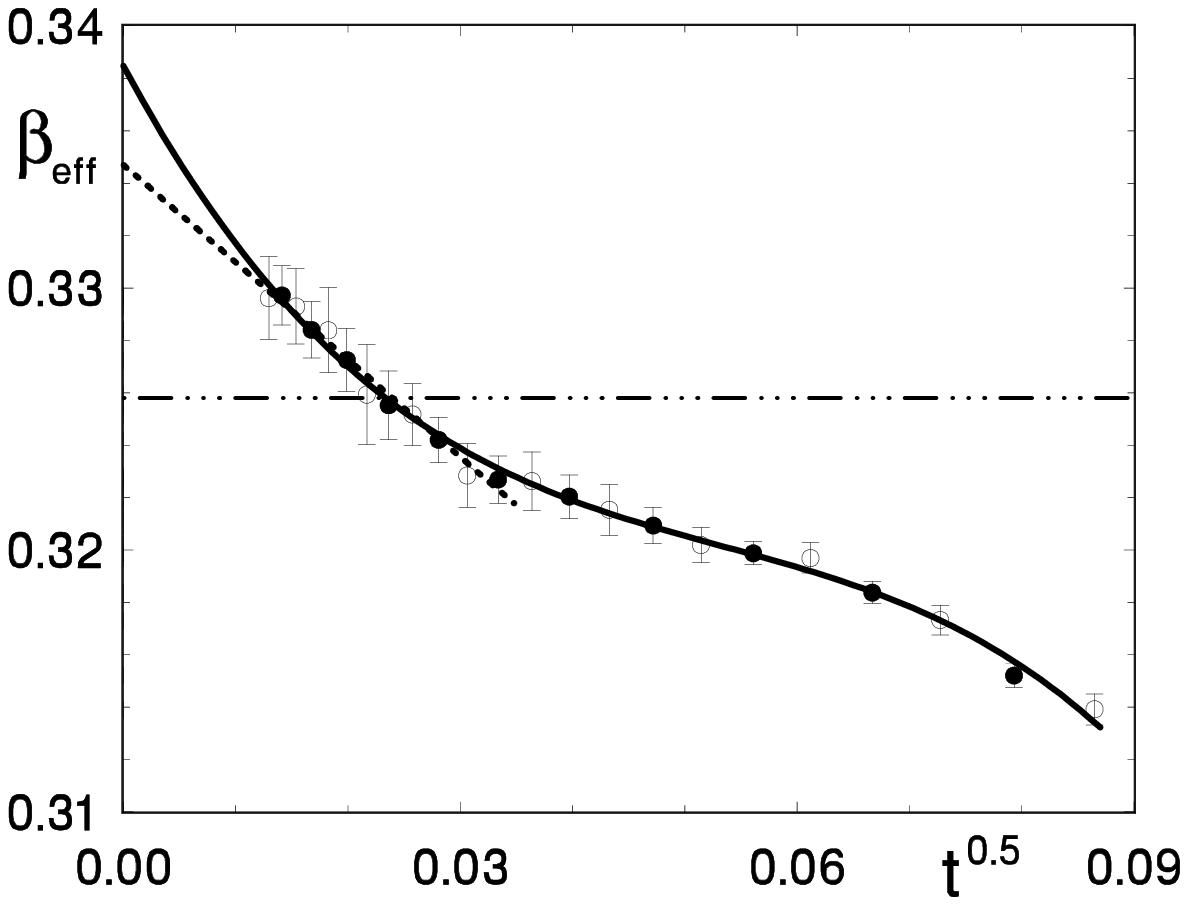,width=8.9cm,height=7.4cm}
        \hspace*{-11ex}
            \psfig{figure=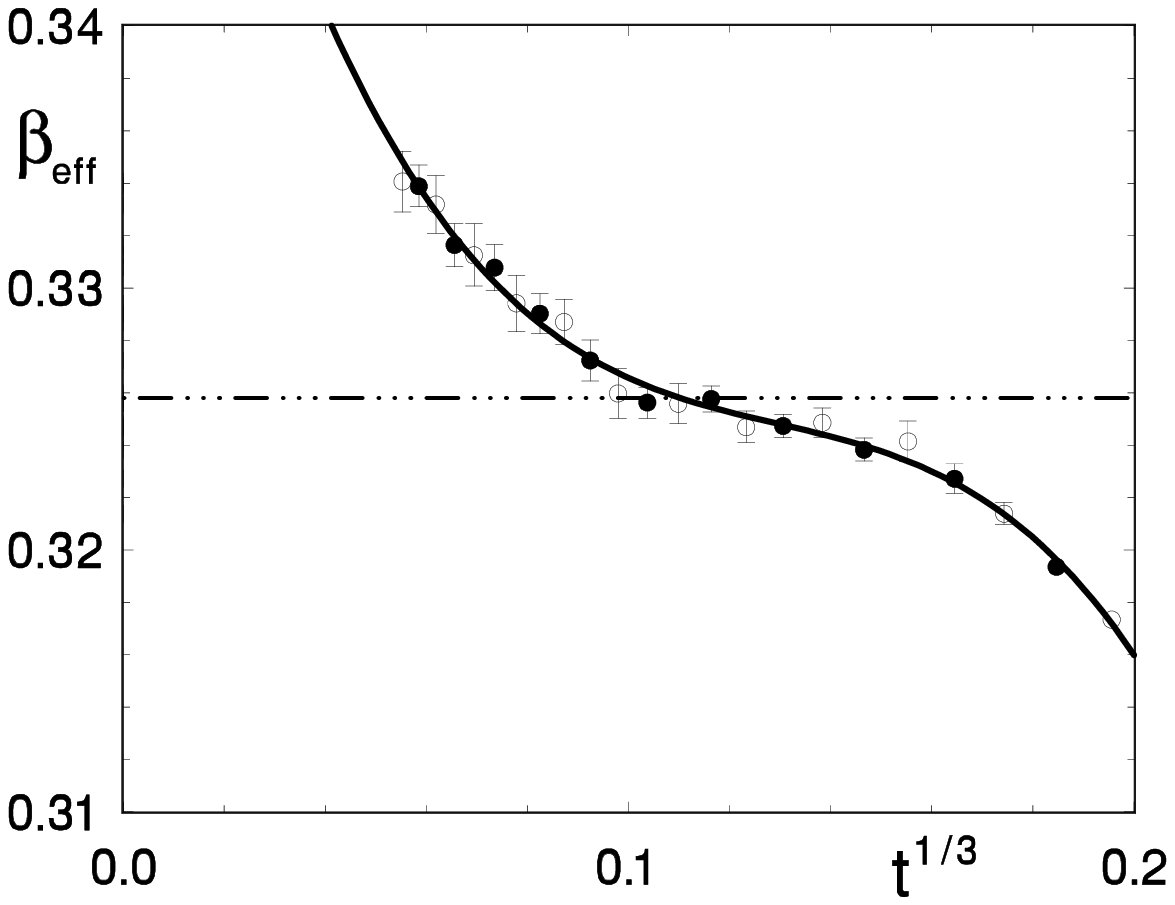,width=8.9cm,height=7.4cm}}
\vspace{-8.5ex}
\caption{\small The effective critical exponent
$\beta_{eff}$  vs $t^{\theta}$ with $\theta=0.5$ (left) and
$\theta = 1/3$ (right).
The values calculated from~(\ref{eq:findif})
with $\nu=0.63$, $K_c=0.22165395$ (left) and $\nu=2/3$, $K_c=0.22165386$
(right) are shown by solid circles (averaged over $K=K_{i+j},K_{i+5-j}$
with $j=0,1,2$) and empty circles (averaged over $K=K_{i+j},K_{i+4-j}$
with $j=0,1$). Solid lines represent the cubic fits, whereas the
dashed line (left) shows the linear $12$--point fit. The dot--dot--dashed
line shows the RG prediction $\beta \simeq 0.3258$~\cite{Justin1}.}
\label{beta}
\end{figure}
We have calculated from~(\ref{eq:findif}) and plotted in Fig.~\ref{beta}
the effective critical exponent $\beta_{eff}$ as a function of 
$t^{\theta}$ with the RG exponents $\nu=0.63$ and $\theta=0.5$ (left), 
as well as with the exponents of GFD theory $\nu=2/3$ and $\theta=1/3$ (right).
The corresponding self consistent estimates of the critical coupling
$K_c=0.22165395(46)$ and $K_c=0.22165386(51)$ have been used, obtained
in Sec.~\ref{sec:critp} from the Binder cumulant data within $L \in [96;384]$.
The results of a weighted averaging
over the estimates obtained at $K=K_{i+j},K_{i+5-j}$ with $j=0,1,2$
are shown by solid circles, whereas the values averaged over
$K=K_{i+j},K_{i+4-j}$ with $j=0,1$ are depicted by empty circles. 
The $\beta_{eff}$ values at $K=K_j, K_{j+\ell}$ have been taken with the weights 
$\propto \ell^2$, which approximately minimize the resulting statistical 
errors, taking into accont that the individual errors are roughly proportional
to $1/\ell$.

In both cases (left and right) the plot of the
effective exponent has an inflection point and is well described
by a cubic curve, although the last 12 points (smallest $t$ values)
can be well fit by a straight line, too.
The linear dashed--line fit with the RG exponents (left) yields
$\beta = 0.3347(24)$ at a fixed
$K_c=0.22165395$. Taking into account the uncertainty in $K_c$,
we have $\beta = 0.3347(52)$. This value reduces to $0.3302(37)$
if we take $K_c=0.2216544(3)$ estimated in~\cite{TB}.
Nevertheless, in both cases it is slightly
larger than the RG value $0.3258$~\cite{Justin1} (dot--dashed line),
supported by the high temperature (HT) series expansion~\cite{BC2},
and also a bit larger than the value of~\cite{TB}
$\beta = 0.3269(6)$ [ansatz~(\ref{eq:TBf})].
The linear fit, in fact, takes into account the leading correction
to scaling only. The cubic fit at $K_c=0.22165395(46)$ (solid curve),
which includes also two sub--leading corrections, tends
to deviate up to $\beta = 0.3385(73)$ ($0.3385(37)$ at a fixed $K_c$)
in a worse agreement with the RG prediction.
Assuming the critical coupling $K_c=0.2216544(3)$ of~\cite{TB},
the cubic fit gives $\beta = 0.3324(52)$.
It is worthy to mention that the fits supporting
the RG value with a striking accuracy can be produced easily
without simulations so close to the critical point.
%The only thing which is necessary is a wish to do this.
For instance, omitting the 5 smallest $K$ values
(10 points on the $\beta_{eff}$ plot),
which roughly corresponds to the simulation range $t > 0.0005$,
the linear $10$--point fit yields $\beta=0.3262(16)$ at $K_c=0.2216544(3)$
(the $K_c$ value of~\cite{TB}) and $\beta=0.3259(15)$ at $K_c=0.22165459(10)$
(the $K_c$ value of~\cite{BST99}).

Taking into account that the $\beta_{eff}$ data are correlated, 
the statistical errors of the extrapolated values have been estimated
as $\left( \sum_i \delta_i^2 \right)^{1/2}$, where $\delta_i$ is the
partial error due to the uncertainty in the $i$--th value of
$\langle M^2 \rangle$.

A self consistent extrapolation within the GFD theory is illustrated in
the right hand side picture (Fig.~\ref{beta}).
The cubic fit gives $\beta=0.366(16)$ in agreement with the expected
exact result $0.375$. Unfortunately, there is still a very large
extrapolation gap, so that we cannot make too serious conclusions
herefrom. It is necessary to go even much closer to the critical
point in this case with simultaneous reduction of the error in
the estimated $K_c$ value.

\section{Estimation of the singularity of specific heat
in 3D Ising model from the finite--size scaling of MC data}
\label{sec:alfa}

It is commonly believed~\cite{Justin1} that the specific heat
$C_V$ of 3D Ising model on an infinitely large lattice has a
power--like singularity, i.~e., $C_V \sim t^{-\alpha}$ at $t \to 0$.
According to the finite--size scaling theory, it would mean that
\begin{equation} \label{eq:CV}
C_V \sim t^{-\alpha} f \left(L^{1/\nu} t \right)
\equiv L^{\alpha/\nu} \tilde f \left(L^{1/\nu} t \right)
\end{equation}
holds at small $t$ in the finite--size scaling region $t \sim L^{-1/\nu}$.
Here $\nu$ is the exponent of correlation length, whereas
$f(z)$ and $\tilde f(z)=z^{-\alpha} f(z)$ are the scaling functions.
The maximum of the $C_V$ vs $t$ plot is located at a certain value of
the scaling argument $z=L^{1/\nu} t$ at $L \to \infty$, which would
mean that the maximum values scale as
\begin{equation}
C_V^{\mbox{\scriptsize max}}(L) \propto L^{\alpha/\nu}
\hspace{3ex} \mbox{at} \hspace{2ex} L \to \infty \;.
\end{equation}
An estimation of the exponent $\alpha/\nu$ from the slope of the
log--log plot then gives us the effective
exponent $(\alpha/\nu)_{eff}= \partial \ln
C_V^{\mbox{\scriptsize max}}(L) / \partial \ln L$, which is varied
due to the corrections to scaling like
\begin{equation} \label{eq:efexp}
(\alpha/\nu)_{eff} \simeq (\alpha/ \nu) + const \cdot  L^{-\omega} \;, 
\end{equation}
where $\omega = \theta/\nu$ is the correction--to--scaling exponent.
Note that specific heat contains also an analytic 
background contibution which influences this behaviour.  
We have found, however, that this influence is very small in the actually 
considered range of sizes if the constant background term is of order 
one, as expected from physically--intuitive considerations.

We allow a possibility that the specific heat has a logarithmic
singularity, as consistent with~\cite{Tseskis} and~\cite{K1}.
It means that Eq.~(\ref{eq:CV}) is replaced with
\begin{equation} \label{eq:CV1}
C_V \sim \ln t \cdot f \left(L^{1/\nu} t \right)
\end{equation}
and~(\ref{eq:efexp}) -- with
\begin{equation} \label{eq:efexp1}
(\alpha/\nu)_{eff} \simeq \frac{1}{\ln (L/L_0)} \;,
\end{equation}
where $L_0$ is a constant length scale.
Although one believes usually that the singularity of $C_V$ is
power--like with $\alpha \simeq 0.11$, no strong numerical evidences
exist which could rule out the logarithmic
singularity~(\ref{eq:CV1}). The problem is that $\ln t$ behaves
almost like a weak power of $t$ with the effective exponent $1/\ln t$,
e.~g., like $t^{-0.11}$ at $t \sim 10^{-4}$.
Moreover, below we will show that the finite--size scaling
of the maximal values of $C_V$ is even very well consistent
with~(\ref{eq:efexp1}) in favour of the logarithmic singularity.

\begin{figure} 
\centerline{\psfig{figure=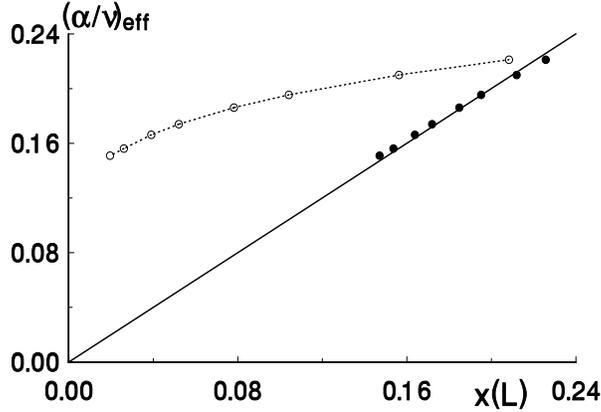,width=9cm,height=7cm}}
\vspace{-8.5ex}
\caption{\small The effective exponent 
$(\alpha/\nu)_{eff}$ of
2D Ising model depending on $x(L)=1/\ln(L/L_0)$ with $L_0=0.572$
(solid circles) and $x(L)=10L^{-1}$ (empty circles). The solid
straight line
represents the approximation $(\alpha/\nu)_{eff}=1/\ln(L/L_0)$.}
\label{alf2d}
\end{figure}

We have tested our method in 2D Ising model,
where the logarithmic singularity of specific heat is well known.
We have considered the effective exponent
\begin{equation} \label{eq:alfn}
(\alpha/\nu)_{eff}(L) = \ln \left[ C_V^{\mbox{\scriptsize max}}(2L)
/ C_V^{\mbox{\scriptsize max}}(L/2) \right] / \ln 4 \;,
\end{equation}
which is defined by finite differences of the log--log plot. 
It coincides with\linebreak
$\partial \ln C_V^{\mbox{\scriptsize max}}(L) / \partial \ln L$
at large $L$ where the log--log plot is almost linear.
Based on exact data, extracted from~(\ref{eq:zz}) and~(\ref{eq:qi}),
we have found that the effective exponent $(\alpha/\nu)_{eff}(L)$
within $L \in [48;512]$ is fairly well described by
ansatz~(\ref{eq:efexp1}) with $L_0=0.572$ and remarkably worse 
described by ansatz~(\ref{eq:efexp}).
It is evident from Fig.~\ref{alf2d}, where $(\alpha/\nu)_{eff}$ 
vs $x(L)= 1/ \ln \left( L/L_0 \right)$ plot (solid circles) well
coincides with the theoretical straight line and is much more linear
than the $(\alpha/\nu)_{eff}$ vs $x(L)=10L^{-1}$ plot (empty circles).
Thus, our method allows to distinguish between the logarithmic
singularity and a powerlike singularity including correction
to scaling of the kind $\propto L^{-\omega}$, where $\omega=1$ holds in 2D case.

The specific heat as well as its derivatives with respect
to the coupling constant $\beta$ can be calculated
easily from the Boltzmann's statistics. Thus, omitting an irrelevant
prefactor, the specific heat is given by
\begin{equation} \label{eq:spech}
C_V = N \left( \langle \varepsilon^2 \rangle
- \langle \varepsilon \rangle^2 \right) \;,
\end{equation}
and the derivatives of~(\ref{eq:spech}) are
\begin{eqnarray} 
\frac{\partial C_V}{\partial \beta} &=& N^2
\left( 3 \langle \varepsilon \rangle \langle \varepsilon^2 \rangle
- \langle \varepsilon^3 \rangle -2 \langle \varepsilon \rangle^3
\right) \label{eq:deriv1} \\
\frac{\partial^2 C_V}{\partial \beta^2} &=& N^3
\left( 12 \langle \varepsilon \rangle^2 \langle \varepsilon^2 \rangle
-3 \langle \varepsilon^2 \rangle^2 -4 \langle \varepsilon \rangle
\langle \varepsilon^3 \rangle -6 \langle \varepsilon \rangle^4
+ \langle \varepsilon^4 \rangle  \right) \label{eq:deriv2} \;,
\end{eqnarray}
where $N=L^3$ is the total number of spins and $\varepsilon$
is the energy per spin.
The maximum of specific heat is located at a pseudocritical
coupling $\tilde \beta_c$ which is defined by the
condition $\partial C_V/ \partial \beta = 0$. It can be
found by the Newton's iterations
\begin{equation}
\tilde \beta_c^{(n+1)} = \tilde \beta_c^{(n)}
- \frac{\partial C_V / \partial \beta}{\partial^2 C_V / \partial \beta^2} \;,
\end{equation}
where $\tilde \beta_c^{(m)}$ denotes the $m$--th approximation
of $\tilde \beta_c$, and the derivatives are calculated
from~(\ref{eq:deriv1}) and~(\ref{eq:deriv2}) at
$\beta = \tilde \beta_c^{(n)}$.

We have used the Wolff's algorithm (in 3D case) to estimate these
derivatives in each iteration consisting of either $5 \cdot 10^5$
(at $L \le 48$) or $10^6$ MC steps. Besides, the
first iteration has been used only for equilibration of the system
retaining the initial estimate of $\tilde \beta_c$. After few
iterations $\beta$ ($C_V$) reaches $\tilde \beta_c$
($C_V^{\mbox{\scriptsize max}}$) within the
statistical error and further fluctuates arround this value.
In principle, the fluctuation amplitude can be reduced to
an arbitrarily small value by increasing the number of MC steps
in one iteration.

An obvious advantage of this iterative method is that the maximal
value of $C_V$ can be evaluated in one simulation without
any intermediate analysis.
Omitting first $5$ iterations, the mean values and the standard deviations
have been evaluated by jakknife method~\cite{MC} from the rest $19$
iterations, except the largest sizes $64 \le L \le 128$, where only 4 iterations
(with twice larger number of MC steps) have been discarded and 11 iterations 
have been used for the estimations. In the most of the cases
the first shuffling scheme discussed in Sec.~\ref{sec:random} has been used as a 
source of pseudo--random numbers, and the simulated values have been verified 
by repeating the simulations at $L=24$, 48, and 96 with the G05CAF generator.
The perfect agreement confirms our results. 

\begin{figure} 
\centerline{\psfig{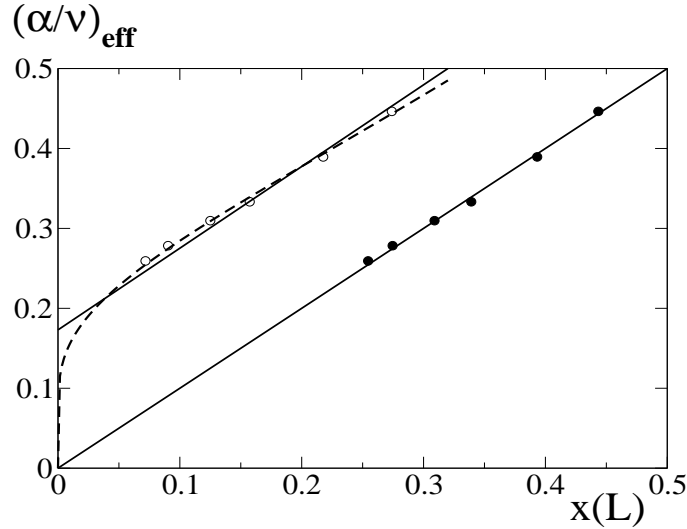}}
%\vspace{-8.5ex}
\caption{\small The effective exponent 
$(\alpha/\nu)_{eff}$
of 3D Ising model depending on $x(L)=1/\ln(L/1.258)$ (solid circles) and
$x(L)=2L^{-0.8}$ (empty circles). The lower solid--line fit represents
the approximation $(\alpha/\nu)_{eff}=1/\ln(L/L_0)$
with $L_0=1.258$, whereas the upper one corresponds to
$(\alpha/\nu)_{eff}=(\alpha/\nu)+2.044L^{-\omega}$ with the
RG exponents $\alpha/\nu=0.173$ and $\omega=0.8$.
The dashed line shows the logarithmic approximation in
the scale of $2L^{-0.8}$.
Statistical errors are within the symbol size. The mean
squared deviations from the lower and upper solid--line
fits are $1.6 \cdot 10^{-5}$ and $8.22 \cdot 10^{-5}$, respectively.}
\label{alfa}
\end{figure}

We have averaged the values of $C_V^{\mbox{\scriptsize max}}$ 
over both simulations at $L=24$, 48, and 96 to reduce the statistical 
errors in the estimated effective critical exponent~(\ref{eq:alfn}).
Our results are summarized in Tab.~\ref{tabula}. 
\begin{table}
\caption{\small The MC estimates of the maximal values of the
specific heat $C_V^{\mbox{\scriptsize max}}$, the pseudocritical couplings
$\tilde \beta_c$, and the effective critical exponents
$(\alpha/\nu)_{eff}$ depending on the system size $L$. The marked
values have been simulated by G05CAF pseudo--random number generator.}
\label{tabula}
\vspace*{2ex}
\begin{center}
\begin{tabular}{|c|l|l|l|}
\hline
\rule[-3mm]{0mm}{7.5mm}
L   & $C_V^{\mbox{\scriptsize max}}$ & $\tilde \beta_c$ & $(\alpha/\nu)_{eff}$ \\ \hline
3   & 16.4445(61)                    & 0.233595(31)     &                      \\
4   & 21.532(10)                     & 0.234207(26)     &                      \\
6   & 28.908(27)                     & 0.231090(33)     & 0.66742(89)          \\
8   & 34.155(28)                     & 0.228561(23)     & 0.5552(14)           \\
12  & 41.481(49)                     & 0.225771(20)     & 0.4464(12)           \\
16  & 46.491(85)                     & 0.224436(13)     & 0.3894(19)           \\
24  & 53.71(10)                      & 0.223207(13)     & 0.3333(19)           \\
24  &$53.65(10)^*$                   &$0.223193(16)^*$  &                      \\
32  & 58.60(15)                      & 0.2226726(96)    & 0.3096(20)           \\  
48  & 65.82(25)                      & 0.2222037(97)    & 0.2784(23)           \\  
48  &$65.86(18)^*$                   &$0.2221990(87)^*$ &                      \\
64  & 71.41(15)                      & 0.2220053(85)    & 0.2593(69)           \\
96  & 78.91(23)                      & 0.2218379(56)    &                      \\ 
96  &$79.01(39)^*$                   &$0.2218387(55)^*$ &                      \\
128 & 83.95(78)                      & 0.2217704(78)    &                      \\ \hline  
\end{tabular}
\end{center}
\end{table}
The effective exponent $(\alpha/\nu)_{eff}(L)$ within $L \in[12;64]$ is rather
well approximated by~(\ref{eq:efexp1}) with $L_0=1.258$, as shown
in Fig.~\ref{alfa} by solid circles and linear 
least--squares fit in the scale of $x(L)=1/\ln (L/L_0)$.
This is a remarkable agreement, taking into account
that ansatz~(\ref{eq:efexp1}) contains only one adjustable
parameter $L_0$. We have tested also ansatz~(\ref{eq:efexp}),
where the exponents $\alpha/\nu=0.173$ and $\omega=0.8$
have been taken from the perturbative RG theory~\cite{Justin1}.
In this case the agreement with the data is worse, i.~e.,
the mean squared deviation is $5.14$ times larger, as shown
in Fig.~\ref{alfa} by empty circles and linear least--squares
fit in the scale of $x(L)=2L^{-0.8}$.
It can be well seen also when comparing the linear fit of circles with
the evidently better dashed--line fit. The latter represents the 
lower straight line in the scale of $2L^{-0.8}$ and shows the expected behavior of  
empty circles at $L \to \infty$ if~(\ref{eq:efexp1}) is the correct ansatz.  
Note that the main deviations of the data points from the fitted lines 
in Fig.~\ref{alfa} are not of statistical character,
since the statistical errors are remarkably smaller than the symbol size
except only the largest $L$ value, where the error is about the symbol size.
Due to this reason we have used simple least--squares approximations,
minimizing the sum of not weighted squared deviations.
These deviations show just the error of the ansatz used and, thus,
indicate that~(\ref{eq:efexp1}) is a better approximation than~(\ref{eq:efexp}) with
fixed exponents $\alpha/\nu =0.173$ and $\omega=0.8$
within the actual range of sizes, at least. 
Moreover, since $L^{-0.8}$ has reached already a rather small value
$0.036$ and, therefore, the second--order correction $\sim L^{-1.6}$
should be very small, the observed deviations can be explained 
easily assuming~(\ref{eq:efexp1}) rather than~(\ref{eq:efexp})
with the RG exponents. 
According to our recent findings (based on unpublished simulation data for $L 
\le 192$) the discrepancy with the RG value $\alpha/\nu =0.173$ can be 
explained by the existence of a negative background contribution $C_0= -25(3)$ 
to $C_V$ which, however, seems to be unphysically large in magnitude.

It is quite possible that the true value of $\alpha=2- d \nu$ 
is remarkably smaller than the RG value $0.11$,
as consistent with our recent results for the exponent $1/\nu$. 
We have estimated $(1/\nu)_{eff}=y_{eff}(L)$ from the derivatives
of the Binder cumulant $U'(L)=dU(L)/ d \beta$ (see Sec.~\ref{sec:critp})
at two system sizes $L/2$ and $2L$.
They should scale like $U'(L) \sim L^{1/\nu}$ at a pseudocritical
coupling corresponding to $U=const$. Our data $U'(16)=-175.34(0.17)$, $U'(32)=-526.65(0.83)$,
$U'(48)=-1007.8(2.1)$, $U'(64)=-1586.8(3.7)$, $U'(96)=-3022.8(8.9)$,
$U'(128)=-4773.6(16.3)$, $U'(192)=-9023.4(48.7)$, $U'(256)=-14284(88)$,
and $U'(384)=-27039(171)$ for $U=1.6$ yield
$y_{eff}(32)=1.5889(18)$, $y_{eff}(64)=1.5901(27)$, $y_{eff}(96)=1.5812(41)$, 
$y_{eff}(128)=1.5851(48)$, and $y_{eff}(192)=1.5805(50)$.
In analogy to the plots in Fig.~\ref{beta}, one may expect an accelerated further deviation
from the perturbative RG value $\simeq 1.587$.

Thus, in spite of the conventional claims that the specific
heat of 3D Ising model has a certain power--like critical singularity,
accurately predicted by the perturbative RG theory,
the actual very accurate MC data for $C_V^{\mbox{\scriptsize max}}$
show that it is even more plausible that the singularity is logarithmic.

\section{Remarks about other numerical results}

There exists a large number of numerical results in the published
literature not discussed here and in~\cite{K1}.
A detailed review of these results is given in~\cite{PV}.
The cited there papers report results which disagree with
the values of the critical exponents we have proposed in~\cite{K1}.

Particularly, the values of the perturbative RG theory
are well confirmed by the HT series expansions~\cite{BC2,AF}.
However, we are somewhat sceptical about such
a support of one perturbative method by another. It could well
happen that the true reason for the agreement is the extrapolative
nature of both methods, according to which both methods describe
a transient behavior of the system far away from the true critical
region. Really, our simulations of the magnetization data
within $t \ge 0.0005$ well confirm the RG value of the
critical exponent $\beta$, whereas the agreement becomes worse
at $t < 0.0005$, where the plot of the effective exponent
shows a remarkable inflection thus indicating that the true critical region,
where the critical exponents can be accurately measured, is
$t \ll 0.0005$. This region, of course, cannot be
directly accessed by the HT series expansions. 
Another problem is that the HT estimation of the critical
exponent $\alpha$~\cite{AF} is based on a priori assumption
that the singularity of specific heat is power--like, whereas
the MC data (Sec.~\ref{sec:alfa}) suggest that it, very likely,
is logarithmic.

According to the finite--size scaling theory, $t L^{1/\nu}$
is a relevant scaling argument, so that not too small values of the
reduced temperature $t$
are related to not too large system sizes $L \sim t^{-\nu}$.
Thus, according to the idea proposed above,
it is quite possible that the MC results for finite systems,
like also the simulations at finite $t$ values,
appear to be in a good agreement with the conventional RG exponents
which are valid within a certain range of $t$ and $L$ values
well accessible in MC simulations.
The huge number of numerical evidences in the published
literature (see~\cite{PV}) for the exponents of the
perturbative RG theory certainly is a serious argument. 
Nevertheless, there is a reason to worry about the validity of
these exponents
at $t \to 0$ and/or $L \to \infty$ because of the following problems.
\begin{itemize}
\item
We have made accurate MC simulations of the magnetization
(Sec.~\ref{sec:mag}) for unusually large system syzes
($L \le 410$) much closer to the critical point
($t \ge 0.000086$ instead of $t \ge 0.0005$)
than in the published literature, and have found that the
agreement with the RG exponent $\beta$ becomes worse in this case.
\item
Our MC estimation of the exponent $1/\nu$, discussed at the end of
Sec.~\ref{sec:alfa}, shows a good agreement with the RG value
$1.587$ at not too large system sizes. However, the agreement becomes 
worse when larger than $L=128$ sizes are included. 
\item
A remarkable deviation of the correction--to--scaling
exponent $\omega$ from the perturbative RG value
$\omega \simeq 0.8$ has been already
reported in literature~\cite{GT}, where also larger than
usually system sizes $L \le 256$
(instead of the conventional $L \le 128$~\cite{HasRev}) have been
simulated in application to the Monte Carlo renormalization
group techniques, yielding $\omega \simeq 0.7$.
\item
The confirmation of RG exponents by MC simulations is not
unambiguous. There exist also examples where the simulation results are
in a remarkable disagreement with these exponents even within the
conventionally considered range of sizes and reduced temperatures.
A particular example is the finite--size scaling of the maximal
values of the specific heat considered in Sec.~\ref{sec:alfa}.
We are afraid that there are also other such examples, but they
are routinely ascribed as unbelievable and do not appear in the
published literature.
\item
It is indeed easy to produce evidences supporting the RG exponents,
as we have shown in Sec.~\ref{sec:mag}, just because these exponents
describe the behavior of a system not too close to criticality,
in the range which can be easily accessed in MC simulations.
The problem is that the usual MC measurements yield only
effective exponents, as shown in Sec.~\ref{sec:test}, which exhibit
quite large variations also in 3D case, as it is particularly well seen
from our plots of the effective exponents.
The leading correction--to--scaling term, included
in the fitted ansatz, also does not completely solve the problem:
in essence it is the same as to make a linear extrapolation
of the effective exponent, but, e.~g., the $\beta_{eff}(t)$
plots in Fig.~\ref{beta} are remarkably nonlinear.
Therefore, only such evidences are really serious, which show very
precisely how the effective exponents provided by simple estimations
converge to certain asymptotic values.
\end{itemize}

If one consider seriously a possibility that the true values
of the critical exponents are those proposed in~\cite{K1}, then a
question arises why the published MC estimates tend to 
deviate greatly from these theoretical values.
In our opinion, the main reason is that
the published simulations have been made too far away from the
true critical region (as regards both $t$ and $L$),
where the critical exponents can be precisely
measured in a simple way routinely used in MC analysis.
The plots of the effective exponents in Fig.~\ref{beta}
provide an evidence for this statement: as we have already
mentioned (Sec.~\ref{sec:test}), simple MC measurements yield just
such effective exponents, and they are varied.
One has to consider corrections to scaling, and not
only the leading one, to get better results.
However, all the existing (MC) correction--to--scaling analyses in
the published literature rely on the RG correction--to--scaling
exponents, therefore the disagreement with the predictions
in~\cite{K1} is not surprising.

Finally, our theory provides a self consistent explanation
why much smaller $t$ values and/or much larger system syzes
$L$ has to be considered, as compared to the known simulations.
It is because the correction--to--scaling exponent
$\omega=\theta/\nu=0.5$
in our theory is remarkably smaller than that of the RG
theory $\omega \simeq 0.8$, which implies that the decay of
corrections to scaling is relatively slow. In fact, the
reduced temperatures we have reached in our simulations of
the magnetization also are still much too large for an accurate
estimation of the critical exponent $\beta$ in the
right--hand--side picture in Fig.~\ref{beta}.
Nevertheless, we can see that the qualitative behavior,
at least, is just such as expected from our theory.

\section{Conclusions}
             
Summarizing the present work we conclude the following:
\begin{enumerate}
\item
Critical exponents and corrections to scaling for different physical
quantities have been discussed in framework of our~\cite{K1} recently 
developed GFD (grouping of Feynman diagrams) 
theory (Sec.~\ref{sec:crex}).

\item
Calculation of the two--point correlation function
of 2D Ising model at the critical point has been made
numerically by exact transfer matrix algorithms 
(Secs.~\ref{sec:algorithm} and~\ref{sec:correc}).
The results for finite lattices including up to
800 spins have shown the existence of a nontrivial
correction to finite--size scaling with a very small amplitude and
exponent about $1/4$, as it can be expected from
our GFD theory.

\item 
Accurate Monte Carlo simulations of the magnetization 
of 3D Ising model have been performed by Wolff's algorithm in the
range of the
reduced temperatures $t \ge 0.000086$ and system sizes $L \le 410$
to evaluate the effective critical exponent $\beta_{eff}(t)$
based on the finite--size scaling.
Estimates extracted from the data relatively far away from the
critical point, within $t \ge 0.0005$, well confirm the 
value $\beta \simeq 0.3258$ of the perturbative RG theory~\cite{Justin1}.
However, the effective exponent tends to increase
above this value when approaching $T_c$.
A self consistent extrapolation does not reveal a contradiction with the 
prediction $\beta=3/8$ of the GFD theory~\cite{K1},
although there is still a large gap between the simulated and 
extrapolated values.
The convergence of $\beta_{eff}$ to the value of GFD theory
$\beta=11/28$ has been observed experimentally in Ni~\cite{SRS},
where the asymptotic value $0.390(3)$ has been found.

\item
An iterative method has been proposed (Sec.~\ref{sec:alfa})
which allows a direct
simulation of the maximal values of the specific heat,
depending on the system size $L$.
The simulated data for 3D Ising model within $6 \le L \le 128$ 
apparently show 
a better agreement with the logarithmic critical singularity
of the specific heat predicted in~\cite{Tseskis}
(and consistent with our result $\alpha=0$) than with the
specific power--like singularity proposed by the perturbative
RG theory~\cite{Justin1}.
%The idea that the critical exponent $\alpha$ is somewhat
%smaller than the RG value $0.11$ is supported also by our MC data
%for the derivative of the Binder cumulant within $L \le 384$
%(end of Sec.~\ref{sec:alfa}).

\end{enumerate}

\section*{Acknowledgements}

This work including numerical calculations of the 2D Ising model 
have been performed during my stay at the Graduiertenkolleg 
\textit{Stark korrelierte
Vielteilchensysteme} of the Physics Department, Rostock University,
Germany.

\end{document}